\documentclass[conference]{IEEEtran}
\IEEEoverridecommandlockouts
\usepackage{cite}
\usepackage{amsmath,amssymb,amsfonts,amsthm}
\usepackage{algorithm}
\usepackage{algpseudocodex}
\usepackage{graphicx}
\usepackage{textcomp}
\usepackage[dvipsnames]{xcolor}
\usepackage{listings}
\usepackage{todonotes}
\usepackage{ragged2e}
\usepackage{caption}
\usepackage{subcaption}
\usepackage{float}
\usepackage{booktabs}
\usepackage{enumitem}
\usepackage{multicol}
\usepackage{multirow}
\usepackage{makecell}
\usepackage{colortbl}
\usepackage[capitalize]{cleveref}
\usepackage{url}
\usepackage{pifont}
\usepackage{mwe}
\usepackage{tabularx}
\usepackage{array}
\usepackage[table]{xcolor}

\usepackage{siunitx}
\usepackage{tikz}
\usetikzlibrary{arrows.meta, positioning, fit, calc, patterns, tikzmark,calc}
\usepackage{dblfloatfix}

\def\BibTeX{{\rm B\kern-.05em{\sc i\kern-.025em b}\kern-.08em
    T\kern-.1667em\lower.7ex\hbox{E}\kern-.125emX}}
\usepackage{macros}

\usepackage[most]{tcolorbox}

\newtcolorbox{HighlightBox}{
enhanced,
boxrule=0pt,frame hidden,
borderline west={4pt}{0pt}{green!70!black},
colback=green!10!white,
sharp corners
}

\tcbset{width=\columnwidth,boxrule=0pt,colback=red,arc=0pt,auto outer arc,left=0pt,right=0pt,boxsep=5pt}

\definecolor{customcolor}{HTML}{9494b8}

\newtcolorbox[auto counter]{ObservationBox}{
    borderline west={1pt}{0pt}{DeepSkyBlue4},
    colback=customcolor!30!white,
    top=1pt,      
    bottom=1pt,   
    left=1pt,     
    right=1pt     
}

\newcommand{\obs}[1]{\begin{ObservationBox} \textbf{Observation~\thetcbcounter:} #1 \end{ObservationBox}}

\newcommand{\dlnetbench}{DLNetBench\xspace}

\newcommand{\bfsixteen}{\texttt{bfloat16}\xspace}

\newcommand{\strdp}{\textsc{DP}\xspace}
\newcommand{\strfsdp}{\textsc{FSDP}\xspace}
\newcommand{\strdppp}{\textsc{DP+PP}\xspace}
\newcommand{\strdppptp}{\textsc{DP+PP+TP}\xspace}
\newcommand{\strdpppep}{\textsc{DP+PP+EP}\xspace}

\newcommand{\world}{\textsc{W}\xspace}
\newcommand{\ir}{\textsc{IR}\xspace}
\newcommand{\er}{\textsc{ER}\xspace}

\newcommand{\iallreduce}{\textsc{IAllReduce}\xspace}
\newcommand{\allreduce}{\textsc{AllReduce}\xspace}
\newcommand{\iallgather}{\textsc{IAllGather}\xspace}
\newcommand{\allgather}{\textsc{AllGather}\xspace}
\newcommand{\reducescatter}{\textsc{ReduceScatter}\xspace}
\newcommand{\alltoall}{\textsc{AllToAll}\xspace}
\newcommand{\sendrecv}{\textsc{SendRecv}\xspace}
\newcommand{\send}{\textsc{Send}\xspace}
\newcommand{\recv}{\textsc{Recv}\xspace}
\newcommand{\waitall}{\textsc{WaitAll}\xspace}
\newcommand{\wait}{\textsc{Wait}\xspace}
\newcommand{\coll}[2]{$#1_{#2}$\xspace}

\newcommand{\intraswitch}{\textsc{Intra-L1}\xspace}
\newcommand{\intragroup}{\textsc{Intra-Group}\xspace}
\newcommand{\intergroup}{\textsc{Inter-Group}\xspace}

\begin{document}

\title{Characterizing the Scalability and Performance of Large-Scale AI Training Under Multi-Tenancy}

\author{
\IEEEauthorblockN{1\textsuperscript{st} Jacopo Raffi}
\IEEEauthorblockA{\textit{University of Trento}\\
jacopo.raffi@unitn.it}
\and
\IEEEauthorblockN{2\textsuperscript{nd} Thomas Pasquali}
\IEEEauthorblockA{\textit{University of Trento}\\
thomas.pasquali@unitn.it}
\and
\IEEEauthorblockN{3\textsuperscript{rd} Lorenzo Piarulli}
\IEEEauthorblockA{\textit{Sapienza University of Rome}\\
piarulli@di.uniroma1.it}
\and
\IEEEauthorblockN{4\textsuperscript{th} Filippo Spiga}
\IEEEauthorblockA{\textit{NVIDIA}\\
fspiga@nvidia.com}
\and
\IEEEauthorblockN{5\textsuperscript{th} Marco Faltelli}
\IEEEauthorblockA{\textit{ENEA}\\
marco.faltelli@enea.it}
\and
\IEEEauthorblockN{6\textsuperscript{th} Andreas Herten}
\IEEEauthorblockA{\textit{Forschungszentrum Jülich GmbH}\\
a.herten@fz-juelich.de}
\and
\IEEEauthorblockN{7\textsuperscript{th} Domenico Siracusa}
\IEEEauthorblockA{\textit{University of Trento}\\
domenico.siracusa@unitn.it}
\and
\IEEEauthorblockN{8\textsuperscript{th} Daniele De Sensi}
\IEEEauthorblockA{\textit{Sapienza University of Rome}\\
desensi@di.uniroma1.it}
\and
\IEEEauthorblockN{9\textsuperscript{th} Flavio Vella}
\IEEEauthorblockA{\textit{University of Trento}\\
flavio.vella@unitn.it}
}

\maketitle
\begin{abstract}
Characterising AI workload performance on modern HPC systems requires understanding both their scalability in isolation and their behaviour under concurrent execution. However, the interplay among parallelisation strategies, network congestion, compute capability, and interconnect technologies remains poorly understood.
This work investigates the performance and scalability of AI models up to \num{2400} GPUs. We quantify the communication overheads and their impact across different interconnects by evaluating scale-up, scale-out, and rack-scale configurations under multiple allocation schemes. 
Finally, we study how multiple concurrent training jobs interfere with each other by designing a realistic noise model.
We design a benchmark suite of AI models to evaluate the performance of five distinct parallelisation strategies across different supercomputing clusters, including Alps, Leonardo, LUMI, JUPITER, NVL72 GB300, and DGX A100. Our work provides a systematic characterization of the scalability and execution efficiency of distributed AI training, while offering key insights into performance behavior under realistic multi-tenant scenarios.
\end{abstract}


\section{Introduction}
The rapid growth of modern AI models has made large-scale distributed training a central challenge for high-performance computing systems. Training workloads now span hundreds to thousands of GPUs, driven by increasingly large models such as GPT-3 and LLaMA ~\cite{brown2020languagemodelsfewshotlearners, grattafiori2024llama3herdmodels}. While compute capabilities have grown rapidly, achieving efficient scaling remains difficult, as communication and network behavior often become the dominant factors in overall performance ~\cite{anthony2024demystifying,wang2023topoopt,zhang2020network}.

Distributed training relies on GPU-centric communication libraries, such as the NVIDIA Collective Communication Library (NCCL)~\cite{NCCL}, the ROCm Communication Collectives Library (RCCL)~\cite{AMDRCCLDevGuide2025}, and the oneAPI Collective Communications Library (oneCCL)~\cite{oneccl_docs_2025}. As models scale, the cost of communication and synchronization increases, often limiting scalability despite abundant compute resources ~\cite{anthony2024demystifying, zhang2020network}. In practice, training performance is determined by the interaction of multiple system and workload characteristics. Parallelization strategies define communication patterns and synchronization points; network topology and interconnect technologies determine how efficiently these patterns are executed; and job placement and co-scheduled workloads influence the level of contention in shared environments.

Prior work has provided important insights into individual aspects of this problem, including communication overheads and scalability limits~\cite{anthony2024demystifying, zhang2020network}, topology-aware optimizations~\cite{wang2023topoopt}, and congestion effects in large-scale systems~\cite{rajasekaran2022congestion}, and fine-grained characterization of power, performance, and hardware utilization on single- and multi-GPU nodes~\cite{elsayed2026characterization}. However, these studies typically consider only a subset of factors, often focusing on a single system, simplified network conditions, or isolated workloads. As a result, there is still limited understanding of how distributed AI training behaves under realistic conditions that combine large-scale, diverse parallelization strategies, complex network topologies, and multi-tenant execution. This limitation is particularly relevant in light of recent studies showing that production workloads and network traffic exhibit significant heterogeneity and complexity, which are not captured by commonly used models and assumptions ~\cite{hu2024characterization,huang2023encore}.

In this paper, we present a systematic study of distributed AI training performance across systems, scales, and architectures. We evaluate platforms ranging from tightly coupled node-level systems to rack-scale designs such as NVIDIA GB300 NVL72, and up to leadership-class supercomputers including Alps, JUPITER, Leonardo, and LUMI. Across these environments, we compare communication libraries from AMD and NVIDIA, and analyze a diverse set of interconnect technologies and topologies, including Dragonfly and Dragonfly+. This breadth allows us to identify performance trends that are consistent across systems, as well as those that are specific to particular architectural choices.

To enable this analysis, we developed DLNetBench, a benchmark framework that reproduces the communication behavior and synchronization structure of modern distributed training workloads. DLNetBench directly issues the collective operations induced by each parallelization strategy, while approximating the computation–communication balance of workloads through hardware-aware timing models. This design enables precise control over job placement, reproducible experiments across systems, and systematic exploration of concurrent workload configurations, which are difficult to achieve with existing deep learning frameworks.

Using this framework, we study how network topology, job placement, and workload concurrency interact to determine the performance of distributed training workloads. Rather than analyzing these factors independently, we focus on their combined effects across systems and scales, and on how they shape both scalability and performance variability in realistic multi-tenant environments.
Our results show that scaling behavior observed under isolated conditions can be misleading, as interconnect characteristics, job placement, and multi-tenant interference can fundamentally reshape performance at scale. By systematically exposing these effects across systems, this work provides practical guidance for designing, deploying, and evaluating large-scale AI training workloads in realistic environments, and helps identify the design choices that matter most for scalable and efficient training in next-generation AI systems and Giga-Factory-scale infrastructures.

\section{Research Questions and Methodology}\label{sec:methodology}

To study distributed AI training under realistic conditions, we design a methodology that models three interacting factors: communication structure, computation time, and network interference. These factors are explicitly controlled to reproduce the execution behaviour of distributed training workloads while enabling systematic comparisons across systems, placements, and configurations.

This methodology is motivated by the needs of two distinct user communities. The first comprises AI researchers and practitioners, who typically view the underlying infrastructure as an abstracted compute resource. From this perspective, the key challenge is understanding how network bottlenecks and communication overheads impact the scalability of different parallelization strategies. The second comprises system designers and operators, for whom the central question is how system-level choices influence application performance. In this setting, job placement, topology, and concurrent job interactions become first-order factors, as they can significantly influence scalability and communication efficiency, depending on both workload characteristics and infrastructure design. The methodology and experiments are designed to answer four main research questions. 

\noindent\textbf{R1.} How do compute capability and network characteristics bound the scalability of various parallelization strategies?

\noindent\textbf{R2.} How does job placement within the network topology impact the scaling efficiency of different parallelization strategies?

\noindent\textbf{R3.} How do system noise and contention affect training performance, and to what extent can placement choice mitigate or amplify these effects? 

\noindent\textbf{R4.} Are these phenomena visible only at large supercomputer scale, or do they already emerge at node-level and rack-scale systems?
 
To address these questions, we model their interaction in a unified experimental setting. 
Communication is not modelled as a generic traffic pattern; instead, it is derived directly from the distributed training strategy, thereby capturing the effects of different parallelization strategies, as these induce distinct collectives, message sizes, communicators, and synchronization points. 
At the same time, computation determines when communication operations are triggered and how they overlap in time. To capture this interaction and maintain a controlled setup, we approximate compute phases using a hardware-aware model based on roofline analysis~\cite{roofline}, preserving the balance between computation and communication phases without relying on full framework execution (e.g., PyTorch). This model calibrates execution delays based on the arithmetic intensity of transformer layers and the theoretical peak throughput and memory bandwidth of the target GPU. 
Finally, network interference is introduced through concurrent job configurations that reflect the shared, multi-tenant nature of modern supercomputing and AI platforms, allowing us to study how contention affects performance under realistic workload conditions.

We consider five primary parallelization strategies that span the design space of modern distributed training. Data Parallelism (\strdp)~\cite{ddp} replicates the model across devices and synchronizes gradients globally. Fully Sharded Data Parallelism (\strfsdp)~\cite{fsdp} reduces memory footprint via hybrid sharding, distributing parameters within a group of devices and materializing them on demand. Combined Data and Pipeline Parallelism (\strdppp)~\cite{dppp} partitions the model sequentially across devices, while combined Data, Pipeline, and Tensor Parallelism (\strdppptp)~\cite{dppptp} combines pipelining with intra-node parameter sharding. Finally, combined Data, Pipeline, and Expert Parallelism (\strdpppep)~\cite{ep} targets Mixture-of-Experts architectures, where tokens are dynamically routed to specialized sub-networks across different GPUs, introducing intensive all-to-all communication phases. Each strategy dictates a distinct set of collective operations. We categorize these communications into two domains. Intra-Replica (\ir) operations handle the data exchanges required to execute a single model instance, such as passing activations between pipeline stages. Extra-Replica (\er) operations synchronize state, typically gradients, across multiple parallel instances of the model. These collectives are detailed in \cref{tab:collectives_communicators}. To ensure realistic communication volumes, we pair each strategy with representative model architectures, such as Vision Transformers (e.g., ViT-H~\cite{vit}) and Large Language Models (e.g., LLaMA~\cite{llama}, Minerva~\cite{minerva}, and Mixtral~\cite{mixtral}), summarized in \cref{tab:strategies_and_models}.

We implement these AI workloads in \dlnetbench, a benchmark framework designed to expose and control the communication behaviour of distributed training workloads. \dlnetbench directly issues the collective operations prescribed by each parallelization strategy, making communication patterns fully observable and reproducible across systems. Unlike wrappers around existing frameworks such as JAX~\cite{jax} or DeepSpeed~\cite{deepspeed}, \dlnetbench enables explicit and fine-grained control over job placement, communication structure, and concurrency, while avoiding framework-specific effects that prevent systematic comparisons.

\dlnetbench does not execute real tensor operations. Instead, it replaces each compute phase with a calibrated sleep whose duration is derived from the roofline performance model~\cite{roofline}. This approach provides a hardware-aware estimate that captures the effective throughput limit imposed by either arithmetic throughput or memory bandwidth, while remaining independent of framework internals whose behaviour can vary across versions and configurations. For each target GPU architecture, we take peak floating-point throughput and peak memory bandwidth for the \bfsixteen data type directly from the vendor datasheet, establishing a single, reproducible baseline that does not depend on empirical microbenchmarks.

Consider a transformer layer with batch size $B$, sequence length $N$, hidden dimension $d$, feed-forward dimension $H$, number of experts $E$, number of active experts $k$, and element size $s$. For element size, we use $s = 2$ for \bfsixteen\, all other parameters are determined by the model in use (see \cref{tab:strategies_and_models}). The per-layer costs are as follows:

{\footnotesize
\[
\begin{aligned}
\mathrm{attn:}\ 
& \mathrm{FLOPs} = 8BNd^2 + 4BN^2d, 
&& \mathrm{Bytes} = 4d^2 s + 2BNd\,s, \\
\mathrm{mlp:}\ 
& \mathrm{FLOPs} = 4BNdHk, 
&& \mathrm{Bytes} = 2dHsE + 2BNd\,s.
\end{aligned}
\]
}

The arithmetic intensity $\mathrm{I} = \mathrm{FLOPs}/\mathrm{Bytes}$ determines whether execution is compute- or memory-bound. The estimated per-layer compute time is $t = \mathrm{FLOPs}/\min\!\bigl(\mathrm{peak\_flops},\; \mathrm{I}\cdot\mathrm{bandwidth}\bigr)$.
The forward time sums $t_{\mathrm{attn}}$ and $t_{\mathrm{mlp}}$ over all $L$ layers; the backward time is approximated as $2\times$ the forward time. Since DLNetBench replaces real computation with a roofline-based timing model, we compared it against reference implementations in production frameworks. For example, on a single LUMI node (8 GPUs) the PyTorch DP implementation is approximately $\approx2.5$ slower end-to-end than DLNetBench under the same configuration. This is expected, since production frameworks may execute many more ML operators that our compute model intentionally abstracts away. Despite this mismatch in end-to-end time, the dominant compute cost (GEMM) and the communication pattern of each parallelisation strategy remain fully controlled and consistent, which is sufficient for our goal of comparable, reproducible results and realistic performance estimation across systems.

Across all parallelization strategies, we use a batch size of 16 and 16 microbatches; for pure data parallelism, gradients are partitioned into 50 buckets; for FSDP, we use 16 sharding units with a sharding factor of 8; and for the hybrid configurations, we use a pipeline parallelism degree of 8, combined with a tensor parallelism degree of 4 or an expert parallelism degree of 8. Here, the parallelism degrees specify the number of processes participating in each respective parallelization dimension, while the number of buckets or units determines the granularity at which gradients or model parameters are communicated.
The suite supports multiple communication backends, including (CUDA-aware) MPI, NCCL~\cite{NCCL}, RCCL~\cite{AMDRCCLDevGuide2025}, and oneCCL~\cite{oneccl_docs_2025}.

To assess scalability, the relevance of network performance, and the impact of multi-tenancy, we conduct two classes of experiments: baseline experiments, in which a single instance of \dlnetbench runs alone, and concurrent experiments, in which multiple instances are co-scheduled. We perform both classes of experiments under two settings: first, in isolation on reserved portions of the system, and second, through the production SLURM queue, reflecting the experience of a typical user. In the reserved setting, each experiment is allocated only a portion of the system (3 Dragonfly groups), and the resulting network traffic is generated by \dlnetbench instances. In contrast, when jobs are submitted through the regular SLURM queue, resource allocations are determined by SLURM, and the network is shared with traffic generated by workloads from other users.

Concurrent configurations vary in both workload mix and scale, enabling us to emulate realistic contention scenarios based on the heavily skewed workload distributions observed in production AI clusters~\cite{hu2024characterization, jeon2019philly}, where a small number of massive pre-training jobs coexist with numerous smaller allocations. We treat each co-scheduled job simultaneously as a potential source of congestion and as a subject of measurement, departing from the conventional aggressor--victim dichotomy~\cite{chunduri2019gpcnet,desensi2020slingshot} to capture the bidirectional interference inherent in shared AI infrastructure.

Experiments construction, job placement strategies, measurements, and metrics are detailed in \cref{sec:experimental_setup}. \dlnetbench and all related scripts are publicly available at: \url{https://github.com/HicrestLaboratory/DLNetBenchSC26/tree/main}

\begin{table}[t]
    \centering
    \begin{tabular}{lll}
    \toprule
    \textbf{Strategy} & \textbf{Collective}$_{\textbf{scope}}$ & \textbf{Per-rank msg size} \\
    \toprule
    \strdp
      & \coll{\iallreduce}{\world}
      & 25.3\,MB \\
    \midrule
    \strfsdp
      & \coll{\iallgather}{\ir}     & 125.5\,MB\,/\,115.6\,MB \\
      & \coll{\reducescatter}{\ir}  & 125.5\,MB\,/\,115.6\,MB \\
      & \coll{\iallreduce}{\er}      & 125.5\,MB\,/\,115.6\,MB \\
    \midrule
    \strdppp
      & \coll{\allreduce}{\er}  & 2.008\,GB\,/\,1.850\,GB \\
      & \coll{\sendrecv}{\ir}   & 67.1\,MB\,/\,33.6\,MB \\
    \midrule
    \strdppptp
      & \coll{\allreduce}{\ir}  & 33.6\,MB  \\
      & \coll{\allreduce}{\er}  & 4.4\,GB   \\
      & \coll{\sendrecv}{\ir}   & 134.2\,MB \\
    \midrule
    \strdpppep
      & \coll{\allreduce}{\er}  & 905.3\,MB \\
      & \coll{\allreduce}{\ir}  & 200.7\,MB \\
      & \coll{\alltoall}{\ir}   & 8.0\,MB   \\
      & \coll{\sendrecv}{\ir}   & 33.6\,MB  \\
    \bottomrule
    \end{tabular}
    \caption{Collectives, communicator scopes, and per-rank message sizes for each
    parallelism strategy. For \strfsdp and \strdppp, sizes are reported as
    Llama3-8B\,/\,Minerva-7B. Subscripts denote the communicator scope:
    W~(world, all ranks), IR~(intra-replica, within a model replica),
    and ER~(extra-replica, across model replicas).}
    \label{tab:collectives_communicators}
\end{table}

\begin{table}[t]
    \centering
    \begin{tabular}{ll}
        \hline
        \textbf{Strategy}          & \textbf{Model(s)} \\ \hline
        \strdp~\cite{ddp} (\textsc{D})             & ViT-H~\cite{vit} (ViT-H)             \\
        \strfsdp~\cite{fsdp} (\textsc{FSDP})       & LLaMA3-8B~\cite{llama}, Minerva-7B~\cite{minerva} (LaM-8)      \\
        \strdppp~\cite{dppp} (\textsc{DP})        & LLaMA3-8B~\cite{llama}, Minerva-7B~\cite{minerva} (Minv)      \\
        \strdppptp~\cite{dppptp} (\textsc{DPT})  & LLaMA3-70B~\cite{llama} (LaM-70)     \\
        \strdpppep~\cite{ep} (\textsc{DPE})      & Mixtral-8x7B~\cite{mixtral} (Mxt)    \\ \hline
    \end{tabular}
    \caption{Parallelization strategies and associated models used in \dlnetbench. Between brackets, we report the short name version we will use during results analysis in~\cref{sec:analysis}.}
    \label{tab:strategies_and_models}
\end{table}

\begin{table*}
\caption{Comparison of supercomputers used in this manuscript. Between brackets, we report the TOP500~\cite{top500} rank.}
\label{tab:systems}
\centering
\scriptsize
\renewcommand{\arraystretch}{1.0}
\newcolumntype{Y}{>{\hsize=1.1\hsize\linewidth=\hsize}X}
\begin{tabularx}{\textwidth}{%
    >{\hsize=.45\hsize\linewidth=\hsize}X Y Y Y Y Y%
    }
\toprule

& \textbf{Alps} (\#8) &\textbf{JUPITER} (\#4) & \textbf{Leonardo} (\#10) & \textbf{LUMI} (\#9) \\
\midrule
\textbf{CPU} & $4\times$ NVIDIA {Grace} (72 core) & $4\times$ NVIDIA {Grace} (72 core) & Intel \textit{Ice Lake} Xeon 8358 (32 core) & AMD \textit{Trento} EPYC 7A53 (64 core) \\
\textbf{GPU} & $4\times$ NVIDIA Hopper &  $4\times$ NVIDIA Hopper & $4\times$ NVIDIA Ampere A100 & $4\times$ AMD MI250X (8 GCDs) \\
\textbf{NIC} & $4\times$ HPE Cray \qty{200}{\giga\bit\per\second} Cassini-1 & $4\times$ NVIDIA Mellanox \qty{200}{\giga\bit\per\second} Connect-X7& $2\times$ dual-port NVIDIA Connect-X6 (\qty{100}{\giga\bit\per\second} per port) & $4\times$ HPE Cray \qty{200}{\giga\bit\per\second} Cassini-1 \\
\bfsixteen & %
    \qty{10644}{\peta FLOP \per\second} 
    & \qty{23300}{\peta FLOP \per\second} 
    & \qty{4313}{\peta FLOP \per\second} 
    & \qty{4562}{\peta FLOP \per\second} 
    \\
\textbf{Intra-node\newline Interconnect} & NVLink 4.0, $6\times$ links (\qty{1200}{\giga\bit\per\second}) between any GPU pair & NVLink 4.0, $6\times$ links (\qty{1200}{\giga\bit\per\second}) between any GPU pair & NVLink 3.0, $4\times$ links between any GPU pair (\qty{800}{\giga\bit\per\second}) & Between one and four \qty{400}{\giga\bit\per\second} Infinity Fabric links to other GCDs \\
\textbf{Inter-node\newline Interconnect} & HPE Cray Slingshot 11, Dragonfly topology  & NVIDIA InfiniBand NDR, Dragonfly+ topology (25 groups, each a 2-level fat-tree) & NVIDIA InfiniBand HDR, Dragonfly+ topology (23 groups, each a 2-level fat-tree) & HPE Cray Slingshot 11, Dragonfly topology (24 groups) \\
\textbf{Software\newline Environment} & prgenv-gnu/26.3:v1 uenv: GCC 14.3.0, Cray MPICH 9.1.0, libfabric 2.4.0, CUDA 13.1.1, NCCL 2.29.2-1 (with AWS-OFI-NCCL 1.17.2) & GCC 14.3.0, Open MPI 5.0.8, CUDA 13.0.0, NCCL 2.29 &  GCC 12.2.0, Open MPI 4.1.6, CUDA 12.4, NCCL 2.18.5 & Cray clang 16.0.1, Cray MPICH 8.1.27, libfabric v1.15.2, ROCm 6.3.4, RCCL 2.21.5 (with AWS-OFI-RCCL 1.4) \\
\bottomrule
\end{tabularx}

\end{table*}

\section{Systems Description}\label{sec:systems}
In the following, we describe the main characteristics of the systems used in our experimental campaign. 
For the analysis of scale-up scalability, we consider two node-level systems: an NVIDIA DGX A100 platform and a LUMI-G compute node. For rack-scale experiments, we use the NVIDIA NVL72 GB300 platform. For the supercomputer-level analysis, we focus on Alps, Leonardo, LUMI, and JUPITER. 
The summary of the main systems' characteristics are reported in Table~\ref{tab:systems}.

\subsection{NVIDIA DGX A100}\label{sec:systems:a100}
The NVIDIA DGX A100~\cite{dgxa100, a100tcdoc} is a node-scale 8-GPU platform designed for AI workloads. It integrates eight NVIDIA A100 Tensor Core GPUs interconnected through third-generation NVLink and six second-generation NVSwitch chips, forming a fully connected intra-node GPU fabric. Each A100 GPU exposes twelve NVLink connections, providing up to 300~GB/s per direction to the NVSwitch fabric. By terminating GPU links on the NVSwitches rather than directly on peer GPUs, the DGX A100 provides a switched all-to-all topology with uniform communication characteristics across all eight accelerators. As a result, any GPU can communicate with any other at node-local bandwidth, and the system can sustain up to 4.8~TB/s of aggregate bidirectional GPU-to-GPU traffic when all eight GPUs communicate concurrently at peak rate.

\subsection{NVIDIA NVL72 GB300}\label{sec:systems:nvlgb300}
\begin{figure}
    \centering
    \includegraphics[width=.8\linewidth]{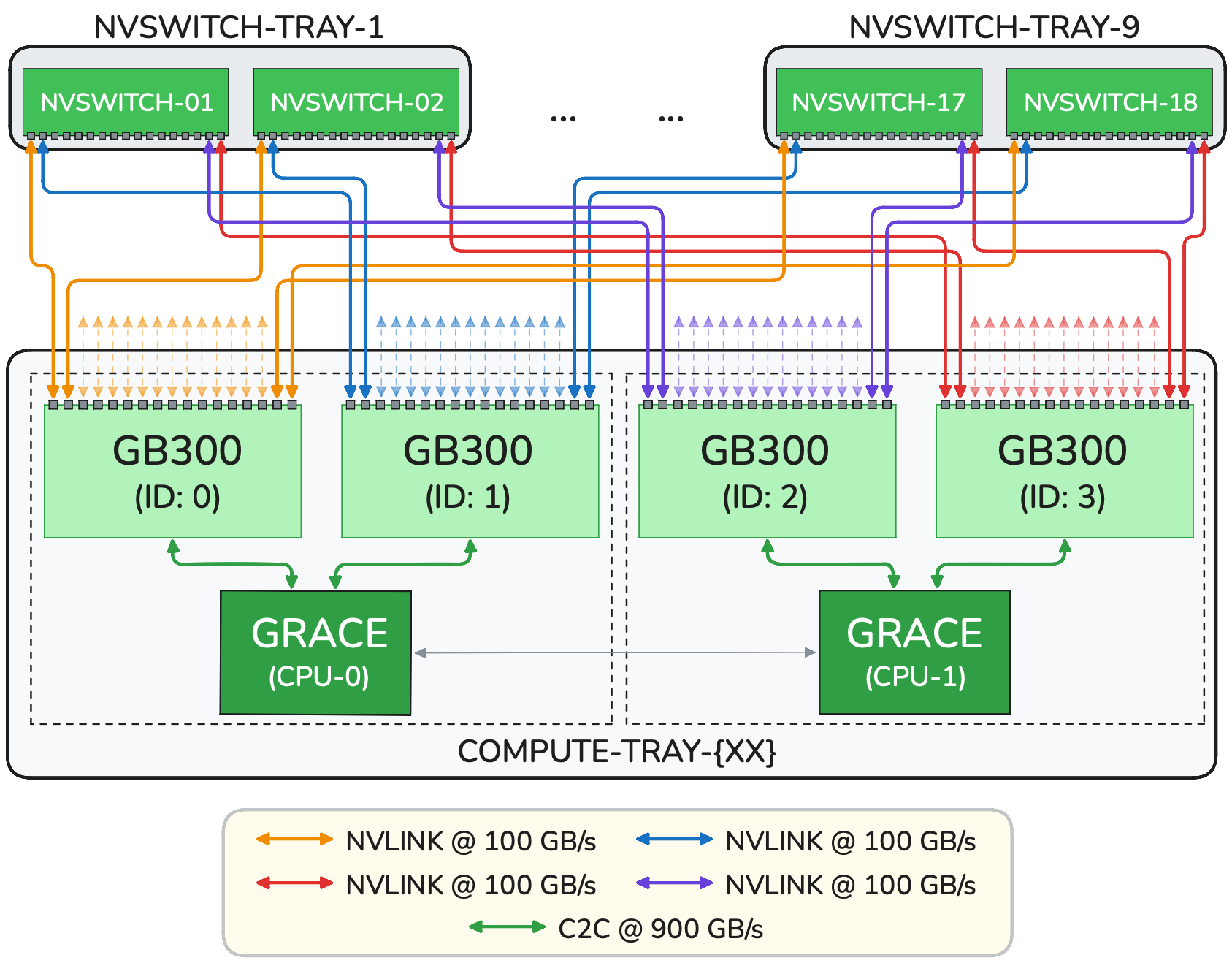}
    \caption{NVIDIA GB300 NVL72 architecture, view of NVLink connectivity from a single compute tray.}
    \label{fig:nvl72}
\end{figure}

For the rack-scale experiments, we consider an NVIDIA GB300 NVL72 system, which integrates 72 Blackwell Ultra GPUs and 36 Grace CPUs, delivering 360~PFLOP/s FP16/BF16, within a single liquid-cooled NVLink domain. The rack is composed of 18 compute trays and 9 NVLink switch trays: each compute tray hosts 2 GB300 Superchips (2 Grace CPUs and 4 GPUs), while each switch tray hosts 2 NVLink switch chips, for a total of 18 switches. Using fifth-generation NVLink and the NVLink Switch System, the rack delivers 130~TB/s of aggregate NVLink bandwidth and full-mesh connectivity across all 72 GPUs. Each GPU exposes 18 NVLink ports, one to each switch chip, providing up to 1.8~TB/s of low-latency GPU-to-GPU bandwidth (\cref{fig:nvl72}). As a result, all GPU communication remains within a single rack-local high-bandwidth fabric~\cite{nvidia_dgx_superpod_gb300_ra}.

\subsection{Alps}\label{sec:systems:alps}
Alps is a 434.90 PFlop/s supercomputer deployed by CSCS~\cite{alps}. The system is composed of HPE Cray Supercomputing EX254n blades, each hosting two nodes. 
\textbf{Node architecture}. Each node is composed of four GH200 \textit{Grace Hopper Superchips}~\cite{gh200doc, gracehopper} connected in an all-to-all topology using NVLink 4.0. Six 200 Gb/s links connect each GH200 pair, for a total of 1.2 Tb/s between any GPU pair.\textbf{Inter-node connectivity}. Each node has one HPE Cray Cassini-1 200 Gb/s Network Interface Card (NIC) for each GH200. Nodes are connected in a Dragonfly topology~\cite{dragonfly} through an HPE Cray Slingshot-11 network~\cite{desensi2020slingshot, khorassani2023high}. Each switch has 16 ports to endpoints, 31 ports to other switches within the same Dragonfly group, and 17 ports to switches in other Dragonfly groups.

\subsection{Leonardo}\label{sec:systems:leonardo}
\emph{Leonardo}~\cite{turisini2023leonardo} is a EuroHPC pre-exascale supercomputer owned by the EuroHPC Joint Undertaking and hosted by CINECA. In the November 2025 TOP500 list, it ranks 10th with an HPL performance of 241.2~PFlop/s. We focus on its \emph{Booster} GPU partition, comprising \num{3456} compute nodes.
\textbf{Node architecture}. Each Booster node comprises a single-socket 32-core Intel Xeon\textsuperscript{\textregistered} 8358 CPU and four NVIDIA A100 \emph{Tensor Core} GPUs~\cite{a100tcdoc}, for a total of \num{13824} GPUs. The GPUs are interconnected through NVIDIA NVLink 3.0, while communication with the host CPU and the NIC is provided through PCIe\textsuperscript{\textregistered} Gen4 x16 links.
\textbf{Inter-node connectivity.} Nodes are interconnected by an InfiniBand HDR200 network. Each Booster node is equipped with two dual-port NVIDIA ConnectX-6 adapters, providing four 100~Gb/s network ports, which we model as four independent NICs. The interconnect follows a Dragonfly+ topology, in which each group contains 180 nodes and is organized as a two-level fat tree with 18 leaf and 18 spine switches. Each switch provides 40 ports at 200~Gb/s, configurable as dual 100~Gb/s links. Within a group, each leaf switch connects 10 nodes through 40 100~Gb/s ports and to the spine layer through 18 200~Gb/s uplinks, leaving two 200~Gb/s ports unused; each spine switch connects to the 18 leaf switches and uses the remaining 22 200~Gb/s ports for inter-group links.

\subsection{LUMI}\label{sec:systems:lumi}
LUMI is a 380 PFlop/s supercomputer owned by the EuroHPC Joint Undertaking and hosted by CSC~\cite{lumisupercomputerLUMIsFull}. This paper considers the \textit{LUMI-G} GPU partition consisting of \num{2978} nodes. 
\textbf{Node architecture.} Each node is built around a 64-core AMD EPYC\textsuperscript{TM} 7A53 ``Trento'' CPU with 128~GB of DDR4 memory, and four AMD MI250X GPUs~\cite{amdcdna}. Each MI250X module consists of two Graphics Compute Dies (GCDs), so that a single node contains eight GCDs in total. Every GCD is paired with a 64~GB HBM memory stack, for an aggregate of 128~GB of HBM memory per MI250X module. Because the two GCDs of each MI250X are exposed as separate devices in the software stack, we model a LUMI-G node as an 8-GPU node. From an interconnect perspective, each GCD is attached to a CPU NUMA domain through a 288~Gb/s AMD Infinity Fabric\textsuperscript{TM} link, while pairs of GCDs are interconnected by one to four 400~Gb/s Infinity Fabric links.
\textbf{Inter-node connectivity}. Each MI250X module is connected to a 200 Gb/s Cassini-1 NIC. Nodes are connected in a Dragonfly topology through an HPE Cray Slingshot-11 interconnect composed of 24 groups, with 124 nodes per group. 
There are 31 switches per group, connected in an all-to-all manner. Each switch has 64 ports, 16 connected to the nodes and 48 connected to other switches. 
Each node is connected to two different switches in the same group. Each switch has 16 ports to endpoints, 31 ports to other switches within the same Dragonfly group, and 17 ports to switches in other Dragonfly groups.

\subsection{JUPITER}

JUPITER is a EuroHPC JU supercomputer hosted at Forschungszentrum Jülich, Germany. Currently in build-up~\cite{jupiterbenchmarks}, it is the first exascale system in Europe and ranks \#4 on the TOP500 list, achieving a performance of \qty{1}{\exa FLOP \per \second}. 
\textbf{Node architecture}. JUPITER uses \num{23536} NVIDIA GH200 superchips, a tight combination of a 72-core ARM-based Grace CPU (\qty{120}{\giga\byte} LPDDR5 memory) and a Hopper GPU (\qty{96}{\giga\byte} HBM3 memory), connected with a \qty{900}{\giga\byte\per\second} NVLink C2C bus. Each node features four GH200 superchips, with direct connections between all GPUs (\qty{300}{\giga\byte\per\second} pair-wise, bi-directional bandwidth) and all CPUs (\qty{200}{\giga\byte\per\second} pair-wise, bi-directional bandwidth).
\textbf{Inter-node connectivity}. The nodes of JUPITER are interconnected by an InfiniBand NDR-200/400 network in Dragonfly+ topology. Within a local Dragonfly group, up to 240 nodes are connected in a fat-tree across two levels of switches, 15 lower L1 switches and 16 upper L2 switches. 
A node features four InfiniBand adapters, which connect to L1 switches with \qty{200}{\giga\bit\per\second} bandwidth per link. Between all switches, the link bandwidth is \qty{400}{\giga\bit\per\second}. In total, 25 of these groups form the overall JUPITER topology, with 30 direct links available between each pair of groups, and further bandwidth available through adaptive routing.

\section{Experimental Setup}\label{sec:experimental_setup}
This section describes how we instantiate the methodology from \cref{sec:methodology} into concrete experiments. We define how we construct baseline experiments, concurrent configurations, and how jobs are placed. Finally, we define the terminology and metrics (e.g., slowdown) used throughout the results analysis in \cref{sec:analysis}.

\subsection{Experiment Construction}

\subsubsection{GPU counts per strategy}
Not every strategy is meaningful at every scale. This variation arises from fundamental architectural constraints and convergence trade-offs. For multi-dimensional strategies, a single model replica often spans many devices. For instance, a \strdppptp configuration with 8 pipeline stages and 4 tensor parallel shards requires 32 GPUs just to host one model replica; applying data parallelism on top of this requires scaling in multiples of 32. Similarly, a \strdpppep setup with 8 pipeline stages and 8 experts requires a minimum baseline of 64 GPUs per replica. Conversely, pure Data Parallelism (\strdp) is impractical at massive scales for models like ViT-H. Scaling pure \strdp linearly increases the global batch size, which can severely degrade training convergence. Therefore, we define the feasible GPU counts for each strategy as follows: \strdp $\in \{\textit{2},\,\textit{4},\,\mathbf{8},\,\mathbf{16},\,32,\,64\}$; \strfsdp $\in \{\textit{4},\,\textit{8},\,\mathbf{16},\,\mathbf{32},\,128,\,256\}$; \strdppp $\in \{\textit{4},\,\textit{8},\,\mathbf{16},\,\mathbf{32},\,\mathbf{64},\,128,\,256\}$; \strdppptp $\in \{\mathbf{256},\,\mathbf{512},\,1024\}$; and \strdpppep $\in \{\mathbf{512},\,\mathbf{1024},\,2048\}$. In these sets, \textbf{bold} denotes GPU counts used in concurrent experiments, \textit{italics} indicate rack/node-scale-only runs, and plain text represents baselines only.

\subsubsection{Allocation patterns}\label{sec:experimental_setup_allocation_patterns}
An \emph{allocation pattern} specifies how a system's $G$ GPUs are divided into job-slots $(g_1, \dots, g_k)$ with $\sum_{1\leq j\leq k} g_j \le G$. We generate patterns from two families. \textit{Uniform} patterns assign the same GPU count to every slot: $(g, g, \dots, g)$. The \textit{tier-sampled} patterns draw slot sizes from a weighted tier distribution that explicitly categorizes jobs into three probabilistic tiers: \emph{small} (75\% probability, 8 GPUs), \emph{medium} (20\%, 16, 32, 64, and 512 GPUs), and \emph{large} (5\%, 1024 GPUs). A pattern is constructed by repeatedly sampling a tier, then a GPU count within that tier, and appending the slot if it fits within the remaining budget, stopping once the budget is exhausted. On LUMI, the GPU counts are doubled since it has 8 GPUs per node instead of 4.

\subsubsection{Strategy Assignment and Configuration Diversity}
Given a pattern $(g_1,\dots,g_k)$, a \emph{strategy assignment} $\phi$ maps each job-slot $j$ to a strategy $S_i\in \{$DP$,\dots,$ DP+PP+EP$\}$ with $g_j$ being a feasible GPU count for that strategy. The resulting set of valid assignments is exponentially large, so we sample it in a stratified manner using a \emph{mixture entropy} score that captures how evenly slots are distributed across strategies: $H_\phi = -\sum_{i} \tilde{p}_i \log \tilde{p}_i,\;\tilde{p}_i \propto \sum_{j\,\mid\,\phi(j)=S_i} g_j$.
$H_\phi=0$ when all slots use the same strategy; $H_\phi=\log m$ (with $m$ strategies) when GPU usage is perfectly balanced across strategies. Because evaluating every possible strategy assignment is computationally infeasible, we group the entropy scores, $[0,logm]$, into three bins representing low, medium, and high diversity. By randomly selecting a few configurations from each bin, we can efficiently evaluate the entire spectrum of system behaviours, ranging from isolated strategies to complex, mixed workloads, without needing to test every single permutation.

\subsubsection{Network Topology and Job Placement}\label{sec:placement_classes}

We model each system as a multi-level switching hierarchy:
a \textit{node} groups 4 or 8 GPUs connected all-to-all; an \textit{L1 switch} aggregates typically 10 to 16 nodes; a \textit{group} connects multiple L1 switches (roughly 120 to 240).
Concrete parameters for each system have been given in \cref{sec:systems}. For a job occupying $g$ GPUs, we define its \emph{placement class} based on how the GPUs are distributed across the network hierarchy:
(i) \intraswitch: GPUs span multiple nodes within one L1-switch domain. (ii) \intragroup: GPUs span multiple L1 switches within one group. (iii) \intergroup: GPUs span two or more groups. (iv) We also consider a deep-learning-motivated variant in which GPUs are arranged in blocks of $n$ nodes each pinned to the same L1 switch (\textsc{Intra/Inter-Group-Same-L1-$n$}); the latter are topology-aware allocations that aim at keeping Intra-Replica traffic local. To compare placement classes numerically, we assign each a \emph{locality score} (1 through 3 for the three classes above, with fractional values for the \textsc{Same-L1} variants) that approximates the average network distance between GPUs.

For each strategy assignment, we vary the distribution of job-slots across the system topology. Each job-slot $j$ is assigned a placement class $\kappa_j$. The mean locality score of a configuration, $\bar{s} = \frac{1}{k}\sum_{j=1}^{k} \emph{locality score}(\kappa_j)$,
summarizes its aggregate network locality. Analogously to the entropy bins, we partition the range $[1, 3]$ into three locality bins and draw a small, fixed number of placements per bin per assignment, covering the spectrum from predominantly \intraswitch to predominantly \intergroup configurations.

\subsection{Measurements, Baselines and Slowdown Metric}

Each \dlnetbench application assigns one MPI rank per GPU. Before any measurement is taken, one warmup iteration is performed and excluded from all results to avoid cold-start effects. After the warmup, all \dlnetbench instances perform multiple iterations. The number of iterations has been chosen depending on the parallelization strategy and system to ensure neither too short nor too long runtimes, for instance, \strdp performs from 6 to 10 iterations while \strdpppep from 2 to 4. For each iteration, we collect the total runtime and instrument all communication operations: blocking collectives are timed directly, whereas for non-blocking collectives we record the time spent at the synchronization barrier, capturing the portion of communication that is not hidden behind computation. Even though the timing of a blocking ``send” does not necessarily correspond to the full duration of the underlying data transfer, we still measure and report the time associated with that collective call, as it reflects the overhead experienced by the application.\\
In multi-tenant scenarios, we set a global timeout (around 4 minutes), and we respawn \dlnetbench instances as soon as they finish on the same nodes. We keep the results produced by all runs and iterations (excluding warmups).

\subsubsection{Baselines}
Every valid \textit{(strategy, GPU count, placement)} triple is measured in both isolation (spanning at most 3 Dragonfly groups) and asking SLURM to allocate resources as a normal user would do. The objective is to characterize the per-strategy performance and scalability before any deliberate concurrent workload is introduced.

\subsubsection{Communication Relevance}\label{sec:experimental_setup_comm_relevance}
To characterize the compute-to-communication balance of each workload on each system, we compute for each baseline the fraction of total runtime spent in communication. This ratio serves as a proxy for how network-sensitive a given strategy and model combination is: workloads with a high communication fraction are expected to be more susceptible to network congestion, while those dominated by compute are less so. The communication time per iteration is defined as the sum of all blocking collective durations plus, for non-blocking collectives, the time spent waiting at the synchronization point, that is, the residual communication cost that overlapping computation could not absorb. We call this value \textit{sync\_time}. We compute this metric by first computing all $\textit{sync\_time} / \textit{tot\_time}$ ratios, for each run we take the maximum across ranks, and we then compute the arithmetic mean across runs~\cite{hoefler2015scientific}.

\subsubsection{Throughput}\label{sec:experimental_setup_dlnetbench_throughput}

Training throughput is reported in samples per second, where one sample corresponds to a single training instance (an image or a fixed-length token sequence). Across all parallelization strategies, the total number of samples processed per iteration is simply the global batch size, which is given by the local batch size multiplied by the total number of model replicas. In our setup, each rank reports the total throughput based on its iteration runtime. For each iteration, we select the minimum throughput (that belongs to the rank that finished last)~\cite{hoefler2015scientific}. Then, we gather the throughput of all runs. For the scalability analysis in \cref{sec:analysis}, we report the geometric mean of iteration throughputs.

\subsubsection{Slowdown}\label{sec:experimental_setup_slowdown}
For a job running concurrently with others, slowdown is defined as the ratio of its baseline throughput to its throughput under concurrent execution: $\sigma = \frac{T_{\mathrm{baseline}}}{T_{\mathrm{concurrent}}}$
$\sigma = 1$ indicates no degradation; $(\sigma - 1)\times 100$ is the percentage throughput loss, often referred to as \textit{congestion impact}~\cite{chunduri2019gpcnet}. For a given strategy $S_i$, GPU count $g$, and placement class $\kappa$ within a concurrent configuration, we gather all available throughput measurements according to what was just described in \cref{sec:experimental_setup_dlnetbench_throughput}, then we compute slowdowns using the corresponding baseline throughput (geometric mean of its iterations' throughputs).

\section{Results Analysis}\label{sec:analysis}
This section analyzes the experimental results through the lens of our research questions (\cref{sec:methodology}), connecting observed performance trends to architectural characteristics, parallelization strategies, and placement decisions. 

We secured reservations spanning three Dragonfly(+) groups on all considered large-scale systems. In particular, 2400 GPUs on JUPITER, 1860 on Leonardo, 1320 on Alps (Clariden), and 2720 GPU modules on LUMI. In some cases, some faulty or underperforming nodes have been excluded as indicated by system administrators. The provided figures already consider this. Even within these reserved groups, adaptive routing in production environments may still introduce residual external interference. For our dedicated testbed systems, namely the NVL72 GB300, DGX A100, we omit user-perspective measurements (job submissions on the production SLURM queue), as their low utilization does not provide representative background traffic. 
On each system, we run all baselines and around 15-25 concurrent experiments constructed as described in \cref{sec:experimental_setup}. Collectively, we ran more than 450 experiments for a total of around 5,300 node-hours.
All figures in this section use the short names of parallelization strategies and models provided in \cref{tab:strategies_and_models}.

\subsection{Communication Relevance}\label{sec:analysis_comm_relevance}
\begin{figure*}[t]
    \centering
    \includegraphics[width=\linewidth]{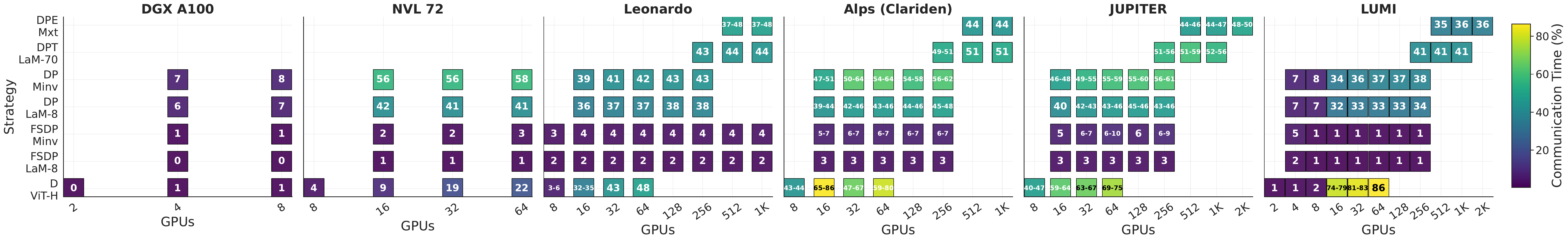}
    \caption{Percentage of runtime spent in communication (detailed in \cref{sec:experimental_setup_comm_relevance})}. Ranges represent the minimum and maximum values obtained with different placement classes (when available and relevant).
    \label{fig:comm_relevance}
\end{figure*}

\cref{fig:comm_relevance} reports the fraction of time spent in communication (detailed in \cref{sec:experimental_setup_comm_relevance}) across all systems and strategies.

On node- and rack-scale systems (DGX A100 and NVL72 GB300), communication overhead is generally limited, although it is not negligible at larger scales. DGX A100 consistently remains below roughly 10\% communication time, indicating that execution is largely compute-bound. NVL72, while still benefiting from high-bandwidth NVLink/NVSwitch interconnects, exhibits higher variability, with communication fractions increasing with scale and reaching up to approximately 41--58\% in some configurations.
On supercomputers, communication becomes a dominant factor even at small scale. With only a few nodes, communication already accounts for roughly 30\% to over 60-80\% of total runtime. This highlights the cost of leaving the node boundary: scale-out communication is significantly slower, leading to both higher absolute communication time and a larger fraction of the total execution.

Across strategies, communication sensitivity varies significantly. \strfsdp consistently exhibits the lowest communication fraction (always below 10\%), demonstrating its ability to effectively overlap communication (\iallgather) with computation. Hybrid strategies (\strdppp, \strdppptp, and \strdpppep) show overall stable communication costs generally in the 30--50\% range, indicating a stronger dependence on network performance. In contrast, \strdp shows the highest communication percentages, with fractions that grow rapidly as the system scales due to its reliance on global \iallreduce operations, often exceeding 70\% of time spent for communication on larger scales.

\obs{These results directly inform R1: communication overhead is minimal intra-node but grows sharply once it crosses nodes, and systems with lower compute-to-network balance (e.g., Leonardo) overlap communication better.}

\begin{figure*}[t]
    \centering
    \includegraphics[width=\linewidth]{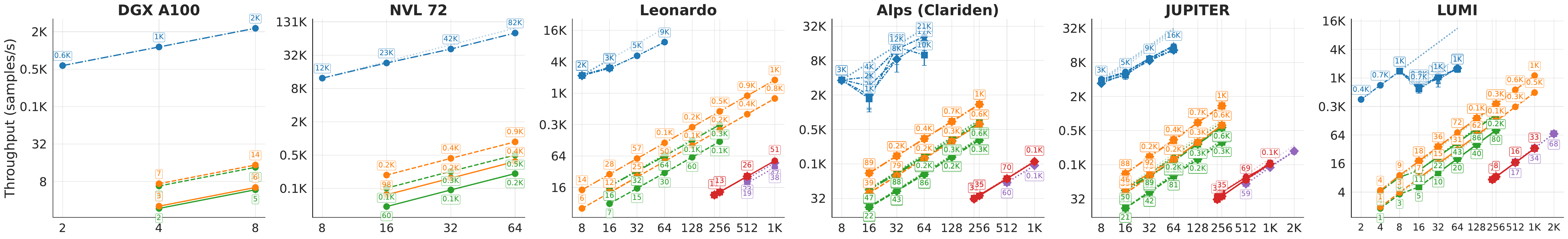}
    \includegraphics[width=\linewidth]{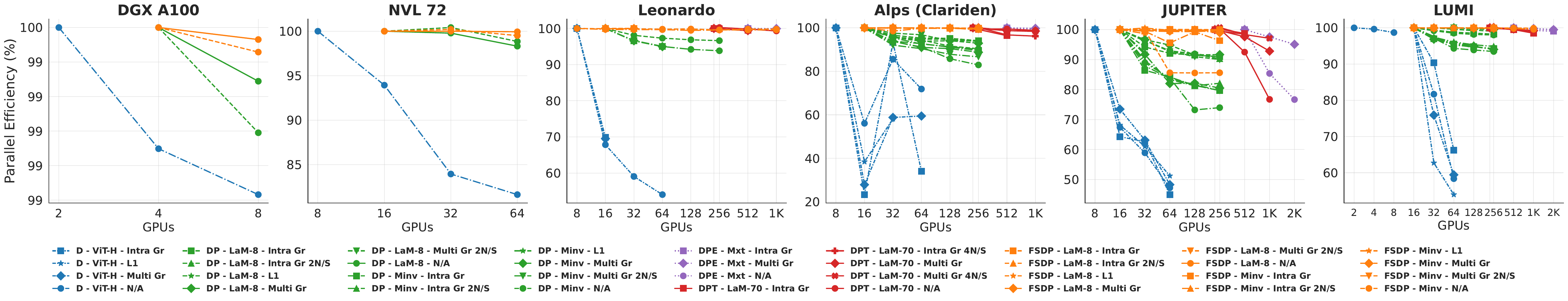}
    \caption{Parallelization strategies scalability and parallel efficiency. Efficiency is computed with respect to the lowest GPU count run; an exception is made for LUMI multi-node experiments, which are normalized on 16 GPUs' throughput}. We report the geometric mean of throughputs computed as explained in \cref{sec:experimental_setup_dlnetbench_throughput}. The dotted lines indicate the ideal scaling. \textbf{The y-axis ranges are fitted to the data} for a better visualization.
    \label{fig:scaling}
\end{figure*}

\subsection{Scalability and Parallel Efficiency}\label{sec:analysis_scalabity}
\cref{fig:scaling} shows that near-ideal scaling is achievable only in regimes where communication remains either negligible or perfectly provisioned.

On the DGX A100, all strategies achieve $\geq$99\% efficiency, confirming that NVSwitch effectively removes communication bottlenecks. The fact that the DGX A100 NVSwitch can perfectly provision communication may be due to the compute-network balance discussed in Observation 1. NVL72 largely preserves this behavior; however, \strdp degrades to 80--85\% at larger GPU counts (32--64 GPUs).
This drop is driven by the increasing number of participants in the \mbox{\iallreduce$_W$}. The efficiency drop is coherent with the 20\% of time spend in communication shown in \cref{fig:comm_relevance}. Regarding \strfsdp and \strdppp, the percentage of time spent in communication remains stable and so does the parallel efficiency. This shows that the high-bandwidth interconnect can perfectly satisfy the communication needs of the hybrid approaches while it is not able to perfectly provision for concurrent \iallreduce. Note that this effect it is probably due to the really high compute capabilities of the Blackwell GPUs, which limit the amount of communication overlapped with compute, thus fully saturating the available bandwidth.

On supercomputers, scalability becomes strongly strategy-dependent. \strfsdp, \strdppp, \strdppptp, and \strdpppep maintain high efficiency, typically above 80--95\%. On JUPITER, the efficiency trends are less stable; in particular, the efficiency of \strdppp, \strdppptp, and \strdpppep drops to around 75\%. \strdp efficiency always degrades significantly. Every time the number of GPUs is doubled, the efficiency drops by 10--30\%, in some cases going below 50\% when running on 64 GPUs. Scalability is fundamentally bounded by the interaction between communication patterns and the interconnect, even with lower compute capabilities. The root cause lies in the communication structure. For pure \strdp on large scales, the communication overhead of global \iallreduce operations increases, reducing the overlap between communication and computation and severely reducing scaling. In contrast, \strfsdp maintains high efficiency because its \iallgather operations achieve good overlap with the compute phase, and its \reducescatter operations are confined to small communication groups (spanning only 8 GPUs, typically across two nodes, or a single node on LUMI). For the hybrid strategies, although \strdppptp introduces more communication overhead compared to \strdppp, its tensor parallel shards are strictly intra-node. Furthermore, the significantly larger compute workload of LLaMA-3-70B, compared to smaller models like LLaMA-3-8B or Minerva-7B, increases the compute-to-communication ratio, reducing the relative impact of these overheads on overall efficiency. \strdpppep benefits from this same favorable ratio. However, while its all-to-all communication is logically restricted to small 8-GPU expert groups, scaling up the total number of GPUs increases the number of model replicas. Because the model is divided into 8 pipeline stages, and each stage operates its own expert communicator, every newly added replica introduces 8 additional independent all-to-all communicators (each fixed at 8 GPUs) to the system. We hypothesize that this multiplication of concurrent transfers might introduce network noise, as these numerous logically independent communicators could begin to disturb one another. This speculation is supported by an observation on JUPITER efficiency in \cref{fig:scaling}. For instance, a \strdpppep configuration localized within 3 Dragonfly groups (purple rhombus) achieves noticeably higher efficiency than a scaled-out configuration spanning 17 groups (purple circle). Similarly, the efficiency drop observed for \strdppptp is for runs spanning 3 and 18 groups.

Absolute throughput is primarily driven by raw compute capability, with systems equipped with more powerful GPUs consistently achieving higher peak values, an ordering that holds clearly across all platforms.
On node- and rack-scale systems, throughput scales smoothly with GPU count and remains close to the expected behaviour, indicating that communication does not significantly limit performance at this scale. On supercomputers, scaling remains broadly consistent with compute capability, but some deviations emerge at larger scale, with higher variability across systems suggesting that communication effects begin to play a role.
A comparison between JUPITER and Leonardo illustrates the dominant role of compute: despite sharing a Dragonfly+ topology, JUPITER achieves around twice the throughput for all parallelization strategies thanks to more powerful GPUs and interconnect bandwidth, while both systems follow similar scaling trends. Comparing JUPITER and Alps instead isolates the effect of the interconnect: with similar compute capability, their throughput is broadly comparable across scales, with no consistent advantage for either system, indicating that the Dragonfly+ topology does not translate into a clear throughput difference between the two. The throughput scaling trend is largely unaffected by placement across all systems (\cref{fig:scaling}), with curves from different placement classes remaining close to one another. Instead, compute capability and interconnect bandwidth are the primary factors shaping throughput. The limited impact of placement arises because communication saturates the links bandwidth, effectively masking the additional latency associated with placements spanning greater network distances (in the scenarios in which baselines are run in isolation).

\obs{Network placement has little to no impact on scaling trends, addressing R2.}

\subsection{Impact of Network Congestion and Multi-Tenancy}

\begin{figure}[t]
    \centering
    \includegraphics[width=\linewidth]{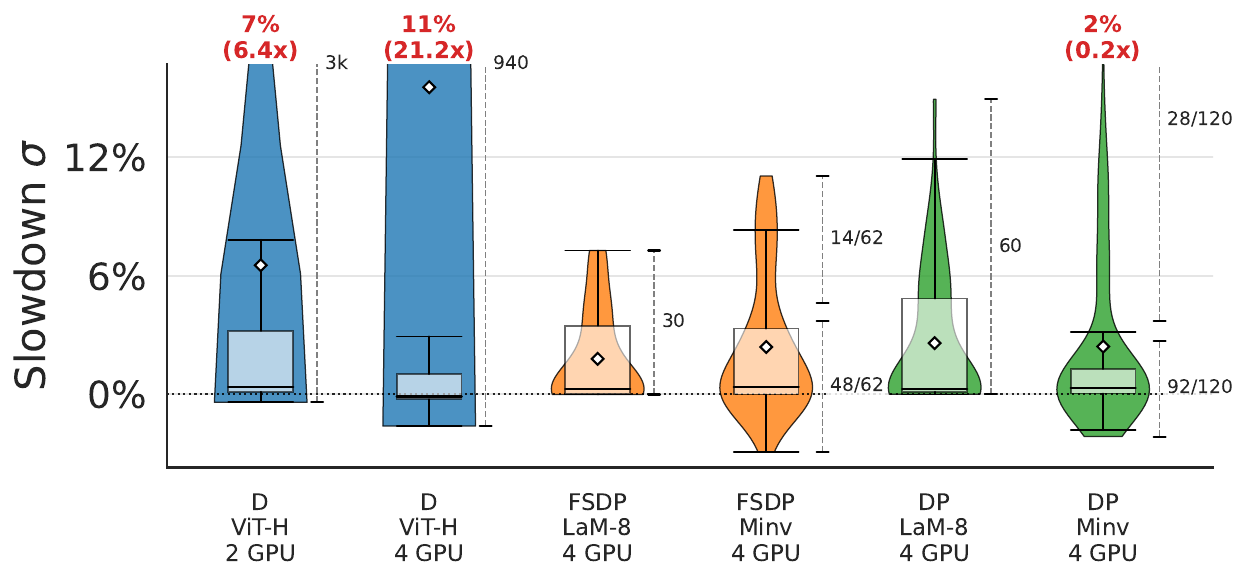}
    \caption{Distribution of slowdown on individual application runs when co-scheduled with other concurrent jobs on DGX A100.}
    \label{fig:cong_plot_dgxa100}
\end{figure}

\begin{figure*}[t]
    \centering
    \includegraphics[width=\linewidth]{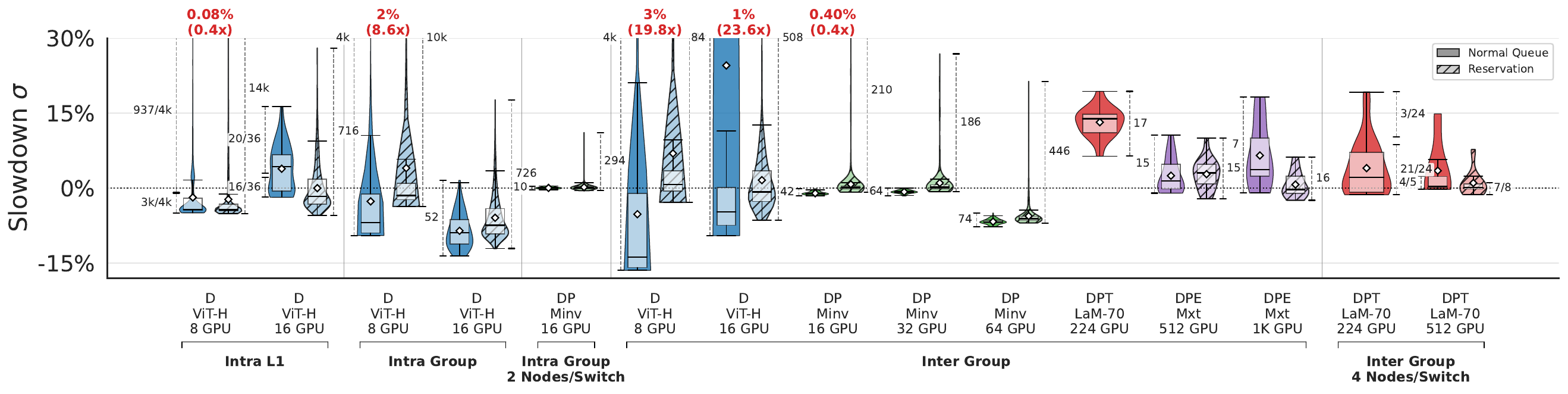}
    \caption{Distribution of slowdown on individual application runs when co-scheduled with other concurrent jobs on JUPITER. Each $x$ tick represents one configuration, defined by its strategy, model, and placement (see \cref{tab:strategies_and_models,sec:placement_classes}).}
    \label{fig:cong_plot_jupiter}
\end{figure*}

\begin{figure}[ht]
    \centering
    \includegraphics[width=\linewidth]{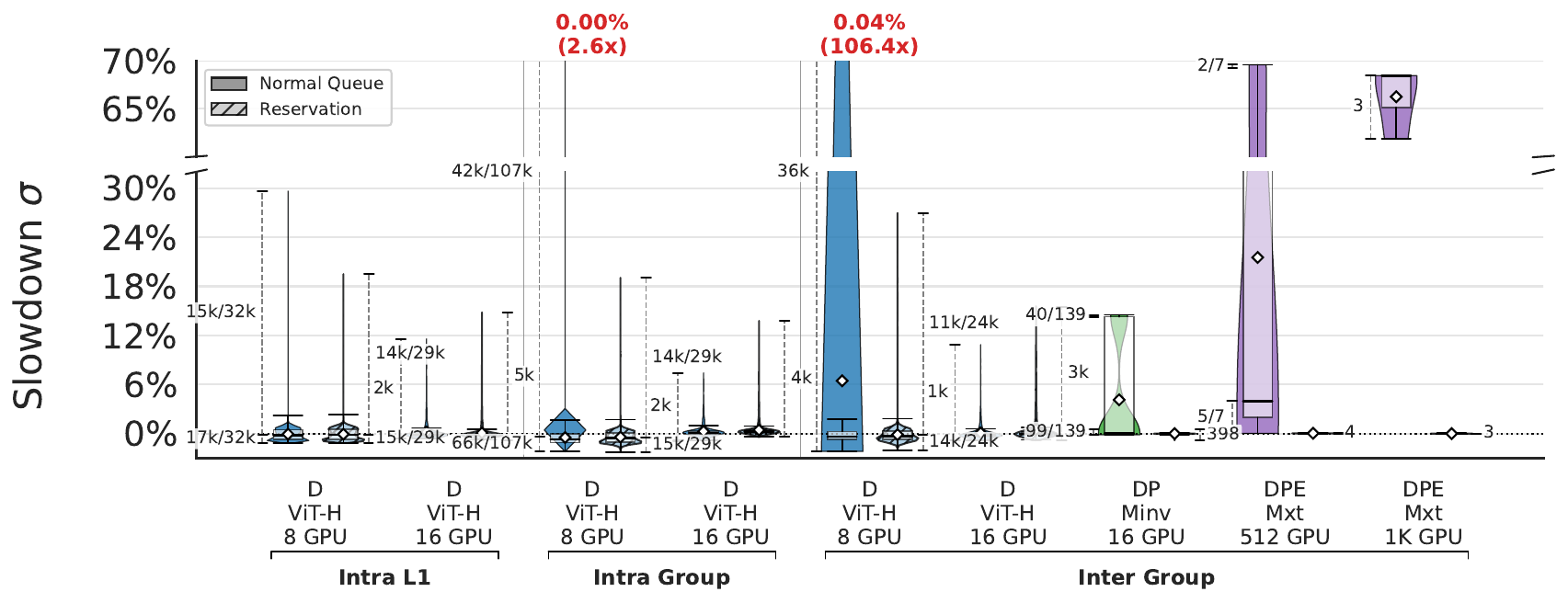}
    \caption{Distribution of slowdown on individual application runs when co-scheduled with other concurrent jobs on Leonardo.}
    \label{fig:cong_plot_leonardo}
\end{figure}

\begin{figure}[t]
    \centering
    \includegraphics[width=\linewidth]{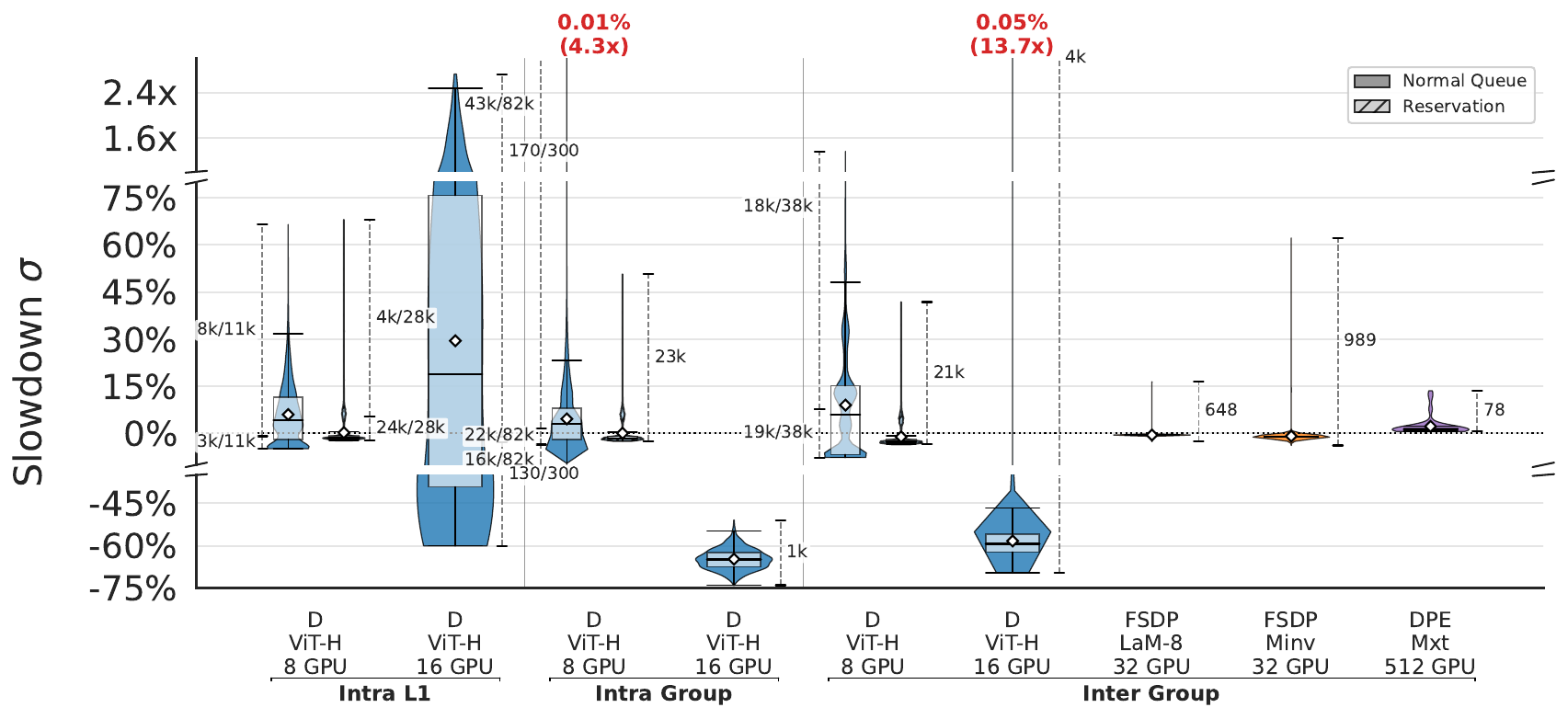}
    \caption{Distribution of slowdown on individual application runs when co-scheduled with other concurrent jobs on Alps.}
    \label{fig:cong_plot_alps}
\end{figure}

\begin{figure}[t]
    \centering
    \includegraphics[width=\linewidth]{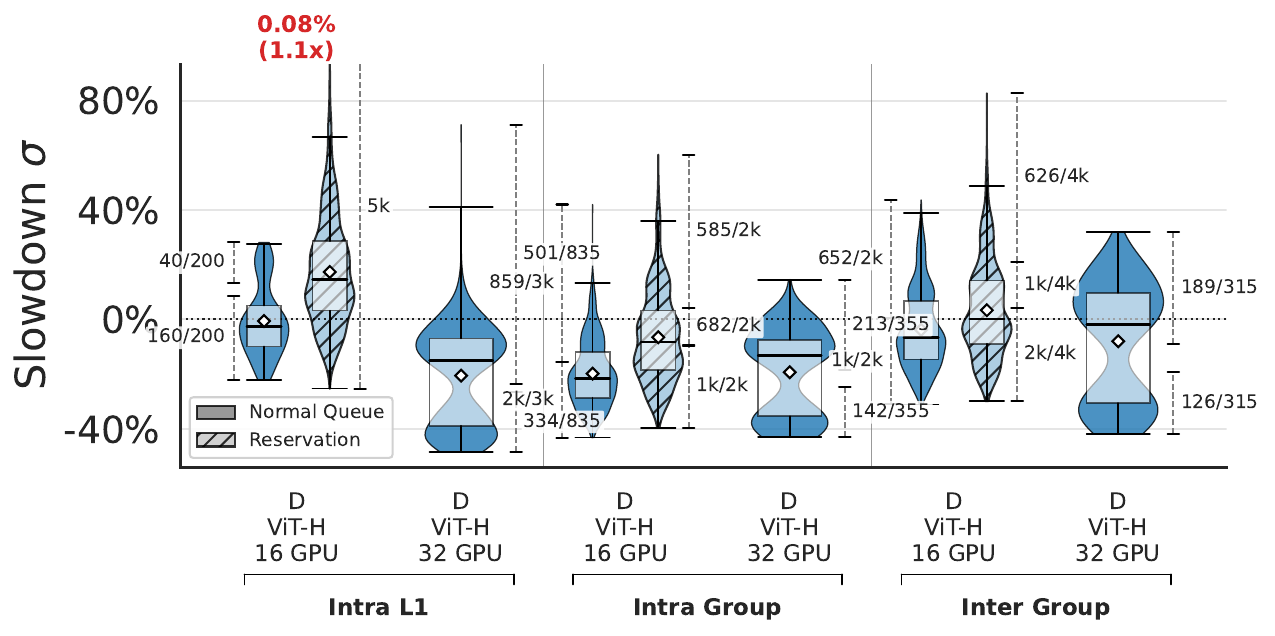}
    \caption{Distribution of slowdown on individual application runs when co-scheduled with other concurrent jobs on LUMI.}
    \label{fig:cong_plot_lumi}
\end{figure}

We now analyze the distribution of slowdown ratios $\sigma$ under co-scheduling. These results expose how contention interacts with communication patterns and job placement. In \cref{fig:cong_plot_alps,fig:cong_plot_lumi,fig:cong_plot_dgxa100,fig:cong_plot_jupiter,fig:cong_plot_leonardo}, each $x$-axis tick corresponds to a specific class of \dlnetbench instances, defined by the combination of parallelization strategy, model, scale (number of GPUs), and placement class (in total we consider 32 distinct combinations). When multiple placement classes are available, they are grouped, as indicated by horizontal brackets below the tick labels. Where reserved and production-queue runs are both available for a given configuration, the corresponding tick is split into two adjacent violins, one per class. The $y$-axis reports the throughput slowdown ratio $\sigma$ (defined in \cref{sec:experimental_setup_slowdown,sec:experimental_setup_dlnetbench_throughput}). For each configuration, we visualize the distribution of slowdowns using boxplots overlaid with violin plots. Boxes represent the interquartile range (IQR, 25th--75th percentile), with the median shown as a solid black line and the mean as a white diamond. Whiskers extend to the most extreme values within $1.5 \times \text{IQR}$.
Dashed brackets alongside each violin partition the samples into disjoint $y$-axis intervals, annotated with the count (or count/total) of runs falling within each. To preserve readability, the $y$-axis is clipped to a fixed range and, for some systems, further split into multiple panels with independent sub-ranges (indicated by diagonal break marks) so that widely separated clusters of slowdown values can each be shown at legible resolution. For configurations with values exceeding the visible range entirely, a red annotation above the plot indicates both the fraction of clipped points and the maximum observed slowdown. Colors denote the parallelization strategy.  Configurations for which no slowdown outside the range $\pm 10\%$ was measured are omitted from the plots; the absence of a given strategy/model/scale/placement configuration therefore indicates that no slowdown was observed for it.

\subsubsection{Scale-Up Systems} On DGX A100 (\cref{fig:cong_plot_dgxa100}), slowdowns are generally below 10--15\%, confirming that intra-node NVSwitch fabrics are not significantly impacted by congestion. However, some extreme outliers (up to $21\times$) indicate that even in tightly coupled systems, transient interference or system noise can disproportionately affect communication-heavy workloads (further details are discussed in \cref{sec:analysis_concurrent_alps}). Negative slowdowns (speedups) are due to transient noise in the measurements.
In contrast, on the NVIDIA NVL72, we do not observe significant slowdowns. All slowdowns are below 1.5\%, with extremely rare exceptions reaching 4\%. Despite its significantly higher compute performance and the corresponding increase in communication demand, the NVSwitch-based fabric consistently sustains all concurrent workloads without observable degradation, effectively eliminating congestion as a performance factor. 

\obs{To answer \textbf{R4}, congestion effects already emerge at node-level scale, though far less severely than at large supercomputer scale: the DGX A100 shows generally low slowdowns with some extreme outliers, whereas the rack-scale NVL72 shows no meaningful slowdown.}

\subsubsection{JUPITER}
In \cref{fig:cong_plot_jupiter}, we observe that when all GPUs are located under the same switch (Intra L1), performance remains mostly unaffected by congestion ($\leq$15\%). While some outliers in 8/16-GPU data parallelism exhibit higher slowdowns, these are likely due to transient system noise rather than network congestion, since intra-switch traffic does not traverse the oversubscribed core fabric. In contrast, intra-group and inter-group placements are sometimes severely impacted, particularly in less regular allocations; for 8-GPU data parallelism, we observe slowdowns reaching $8.6\times$ and $23.6\times$ for intra-group and inter-group cases, respectively. The effects of measurements noise for \strdp are further analyzed in \cref{sec:analysis_concurrent_alps}. The benefit of regular and application-specific allocation is clearly demonstrated by \strdppptp (DPT) on LLaMA3-70B (LaM-70) in 224-GPU runs across different inter-group scenarios. Specifically, in a standard inter-group allocation, executions experience an average slowdown of approximately 15\%. By contrast, using a 4-nodes-per-switch allocation in an inter-group configuration reduces the average slowdown to only 5\%. A similar trend can be observed for \strdppp instances, where 2-nodes-per-switch placements strongly reduce, or even eliminate (violins not shown), the variability in slowdowns. In most cases, runs on the production SLURM queue exhibit greater variance than those executed within 3 reserved groups, as expected: allocated nodes may span multiple Dragonfly+ groups, and network traffic can be affected by other users on the system. Interestingly, this pattern does not hold for \strdppp. In the reservation scenario, approximately 25\% of runs exhibit higher slowdowns. This behavior may be attributed to the strategy's ``pipeline'' peer-to-peer communication pattern, which, under high interconnect utilization, can become vulnerable to delays affecting individual messages along the communication pipeline, resulting in a drop in the overall throughput.

\subsubsection{Leonardo}
The results for the Leonardo supercomputer are reported in \cref{fig:cong_plot_leonardo}. In contrast to the results observed on JUPITER, Leonardo generally exhibits a lower overall impact from network congestion. This difference is largely due to Leonardo's reliance on A100 GPUs, which, being less performant than the H200 used in JUPITER, as discussed in \cref{sec:systems:leonardo}, results in a lower relevance of the communication performance. There are, however, two notable exceptions to this trend. First, data parallelism on 8 GPUs with inter-group allocation shows a noticeably higher variance, with some runs experiencing slowdowns as high as $106\times$. Second, \strdpppep (DPE) at a 1K-GPU scale in the ``no-reservation'' scenario shows a significant performance hit, with average slowdowns reaching 65\%. These runs have been automatically allocated by SLURM on more than 10 Dragonfly+ groups. As more groups are spanned, the probability of being affected by congestion on inter-group links increases. This effect is further exacerbated by the fact that Leonardo's interconnect provides only half the bandwidth of JUPITER's. Moreover, as shown in \cref{fig:comm_relevance}, \strdpppep spends almost half of the runtime time in communication. However, we only observe slowdowns when allocations exceed 3 Dragonfly+ groups.

\begin{figure}[t]
    \centering
    \includegraphics[width=\linewidth]{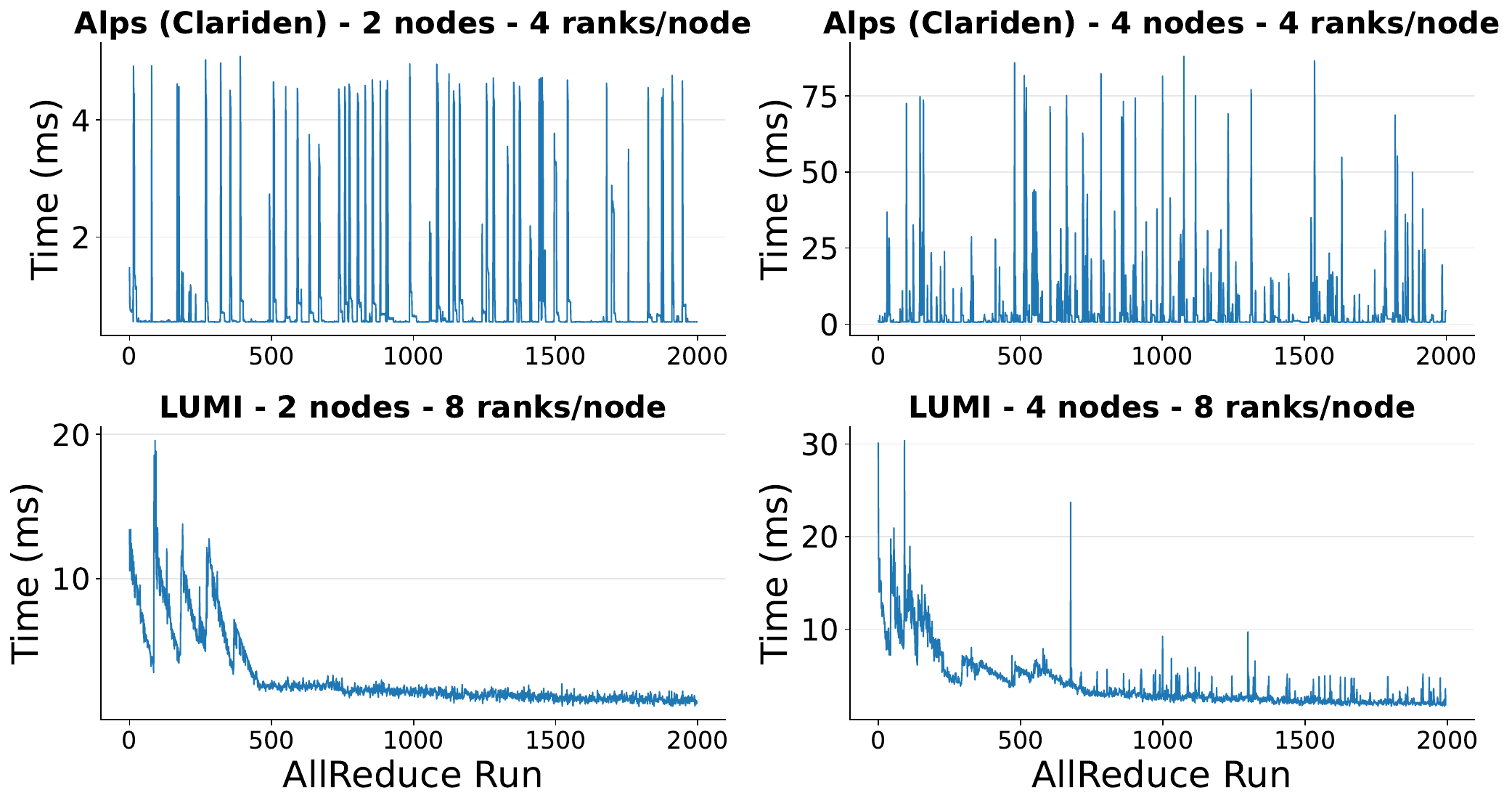}
    \caption{Runtimes of \allreduce collective on Alps and LUMI. Note that $y$ axis is fitted to the data.}
    \label{fig:ar_microbenchmark}
\end{figure}

\subsubsection{Alps}\label{sec:analysis_concurrent_alps}
As becomes evident from \cref{fig:cong_plot_jupiter}, \strdp seems to have high variability in the throughput measurements which reflect on slowdowns directly. Therefore, we analyzed this aspect further. Pure Data Parallelism (\strdp) is an intrinsically simple strategy: each step performs 50 global non-blocking \allreduce operations (one per gradient tensor), followed by a global synchronization on all 50 reductions. Because the underlying model (ViT-H) is comparatively small, per-step compute time is short, leaving little room for communication overhead to be hidden behind computation and making \strdp especially exposed to network-side variability.
To isolate this effect, we ran an \allreduce microbenchmark on Alps and LUMI, using the same message sizes as \dlnetbench \strdp, over each system's native collective backend (NCCL on Alps, RCCL on LUMI). This microbenchmark issues 2000 \allreduce collectives, one at a time, and reports the runtime of the slowest rank. Results are shown in \cref{fig:ar_microbenchmark}. Intra-node performance is stable on both systems, but beyond one node it becomes unstable, with fluctuations amplifying at larger node counts and both systems showing extreme transient outliers. Notably, LUMI's fluctuations decay and stabilize after around 500 runs, whereas Alps continues to show persistent, non-decaying variability.
Because \strdp's per-step runtimes are so small, even minor variance in collective performance translates into large swings in measured throughput. We conjecture this stems partly from rank desynchronization, compounded by our throughput methodology, which uses the last-arriving rank at each step and so is particularly sensitive to any single rank's transient slowdown.
These effects are already visible on DGX A100, JUPITER and Leonardo but worsen on Alps and LUMI.

\cref{fig:cong_plot_alps} shows that \strdp slowdowns variance is consistently high among all scales and placement classes on the Alps system. Accordingly to microbenchmark results, such high variance is due to the variability in runtime of \allreduce communications. Such instability, does not allow to measure a proper baseline, thus, reported absolute slowdowns values are difficult to interpret.

Although \cref{fig:comm_relevance} indicates that communication accounts for only 3--7\% of the total runtime in \strfsdp, this percentage appears sufficient to trigger, in a few isolated cases (8 runs), slowdowns of up to 60\% due to orders of magnitude increase in communication time. Notably, this sensitivity is only observed for Minerva-7B. \strdpppep shows a few $\geq$10\% slowdowns, though exclusively when scaling to 512 GPUs. Note that these slowdowns are only observed in runs on the production queue. Despite \strdppp and \strdppptp spending 40\% to 60\% of their total runtime in communication (\cref{fig:comm_relevance}), they do not seem to be significantly affected by network congestion showing that Alps interconnect can effectively support their communication patterns even in presence of network traffic.

\subsubsection{LUMI}
Similarly to Alps, hybrid strategies spend 30\% to 40\% of their total runtime in communication. Still, \cref{fig:cong_plot_lumi} shows that on LUMI all hybrid strategies exhibit no significant slowdown and are therefore omitted from the figure entirely. This is probably because LUMI features 8 GPUs per node, so the intra-replica communications stays on the scale-up network, and only the extra-replica \allreduce traffic crosses onto the scale-out network. \strdp, in contrast, shows high variance; this is because RCCL requires a substantial number of iterations to stabilize (see \cref{fig:ar_microbenchmark}), and in our experiments it had not yet reached a stable regime, inflating the observed slowdown variance.

System noise and contention affect training performance unevenly, depending on communication pattern and scale, rather than uniformly across all workloads. Congestion is chiefly a scale-out network phenomenon: intra-switch/intra-node traffic is largely unaffected, while cross-group traffic can produce severe slowdowns. Sensitivity is highest for strategies with many small/medium messages and low per-step compute to hide latency (\strdp and \strdpppep) or with chained peer-to-peer dependencies (\strdppp), yet the final impact is strongly system-dependent. System noise, distinct from network contention, also plays a role: unstable collective-library behavior (NCCL/RCCL not yet converged) drives high-variance slowdowns even for intra-switch \strdp traffic. Placement choice is an effective but bounded mitigation. Regular, topology-aware allocation (e.g., fixed nodes-per-switch) can cut average slowdown several-fold or eliminate variability entirely. Reserved allocations generally reduce variance relative to production-queue scheduling, but not universally: pipeline-parallel jobs can show the opposite trend due to their sensitivity to individual message delays. Overall, placement can strongly amplify or suppress contention-driven slowdowns, but its mitigating power depends on the communication pattern. Noise-driven variance from unstable interconnect software remains largely unaffected by placement altogether.

\obs{To answer R3, contention and noise affect performance unevenly across communication patterns and scales; topology-aware placement can mitigate contention up to a certain limit.}

\section{Related Work}\label{sec:related_works}

\subsection{Impact of Network Congestion}\label{sec:general_congestion}
Network congestion caused by concurrent jobs sharing the interconnect fabric has emerged as a critical source of performance variability and a scalability limit. Hoefler et al.~\cite{10.1109/SC.2010.12,5161095} provided an early characterization of how both operating system and network noise degrade parallel application performance, demonstrating that even minor perturbations can create substantial scaling bottlenecks. Similarly, Bhatele et al.~\cite{bhatele2013neighborhood} demonstrated variability of up to $2\times$ for communication-intensive applications, directly correlating these fluctuations with contention from co-running workloads. This landscape is further complicated by the shift toward cloud-based HPC; research indicates that the impact of network noise in cloud environments is even more pronounced than that observed in traditional, on-premise HPC systems~\cite{10.1145/3570609}.

The interaction between network topology and job management also plays a significant role in congestion. Jain et al.~\cite{7013015} explored how different routing strategies and job placement policies affect throughput on Dragonfly networks, highlighting that randomized placement can help mitigate hotspots. Other studies~\cite{desensi2019noise} observed that adaptive routing, while intended to alleviate congestion by using non-minimal paths, can paradoxically generate additional traffic and increase interference for other applications. To address these challenges, specialized tools have been designed to systematically detect and quantify the impact of network congestion: Jha et al.~\cite{jha2020monet} developed Monet to characterize the spatial extent of congestion hotspots, while the GPCNeT~\cite{chunduri2019gpcnet} \emph{aggressor--victim} methodology has become widely used for measuring latency and bandwidth degradation by inducing synthetic contention in shared environments~\cite{desensi2020slingshot,sc24}.

\subsection{Network Contention in AI Clusters}\label{sec:ai_congestion}
In AI-centric datacenters, the impact of congestion is intensified by the bursty, bulk-synchronous traffic patterns inherent to distributed training~\cite{bonato2024fastflow,295483}. Inter-job contention is particularly damaging in GPU-accelerated clusters. The Crux system~\cite{cao2024crux} demonstrated that when a $64$-GPU LLM training job and a $16$-GPU language model share network resources, both GPU utilization and training throughput degrade significantly. The severity of these issues is further explained by the highly skewed nature of AI workloads. Jeon et al.~\cite{jeon2019philly} first identified a power-law distribution in jobs runtimes on Microsoft's Philly cluster, where a small fraction of multi-GPU jobs accounts for the vast majority of GPU time. This observation was reinforced at a larger scale by Hu et al.~\cite{hu2024characterization} in the Acme datacenter, where LLM pretraining jobs represent only $3.2\%$ of the total job count but consume $94.0\%$ of GPU resources, while evaluation jobs constitute $92.9\%$ of the count but use merely $0.8\%$ of resources. 
By integrating these insights on power-law workload distributions with established \emph{aggressor--victim} methodologies, we have developed a new congestion evaluation strategy (Sec.~\ref{sec:experimental_setup}). This approach reflects realistic multi-tenant allocation patterns to faithfully reproduce the conditions observed on production AI supercomputers.

\section{Conclusion and Future Work}\label{sec:conclusion}
This work provides a systematic, large-scale characterization of distributed training performance across six systems spanning node-, rack-, and supercomputer-scale interconnects. Our results show that communication overhead is minimal within a node or rack but becomes highly relevant once workloads cross the node boundary, accounting for 30\% to over 70\% of runtime even at small supercomputer scales. Strategy choice strongly mediates this transition: \strfsdp consistently overlaps communication with computation and sustains high efficiency, pure \strdp degrades sharply with scale due to its reliance on global \iallreduce, and hybrid strategies fall in between, with robustness depending on how their communication groups map onto system topology.

Regarding placement, network placement has little to no impact on aggregate throughput scaling: absolute throughput is governed primarily by compute capability and interconnect bandwidth. This decoupling breaks down under contention.

Our multi-tenancy analysis shows that congestion and system noise affect performance unevenly. Congestion is chiefly a scale-out phenomenon: intra-switch and intra-node traffic remains largely unaffected across all systems, while cross-group traffic can produce severe, highly variable slowdowns. Independent of contention, unstable collective-library convergence (NCCL/RCCL) is itself a source of high-variance slowdowns that placement cannot mitigate. Where congestion is the driver, topology-aware placement is an effective but bounded mitigation: regular, application-specific allocation can cut average slowdowns several-fold, and reserved allocations generally reduce variance relative to production-queue scheduling, though pipeline-parallel strategies can show the opposite trend. At node scale, DGX A100 is not significantly affected by co-scheduling, with only a handful of isolated outliers that are more plausibly attributable to transient system noise than to genuine network congestion, while NVL72's NVSwitch fabric essentially eliminates congestion as a performance factor despite higher compute and communication demands.

Together, these findings indicate that scalable, robust large-scale training performance requires jointly reasoning about parallelization strategy, communication pattern, interconnect characteristics, and topology-aware placement, since no single factor predicts performance under multi-tenancy and in isolation.

Several aspects remain open. Our evaluation uses a limited set of models with fixed parameters, constraining analysis of compute-to-communication balance and achievable overlap. Our Alps/LUMI microbenchmarks suggest collective-library convergence behavior warrants further study to disentangle noise from genuine contention. Future work will broaden models and scales to separate model-specific from strategy-level trends, vary \dlnetbench parameters to probe compute-communication balance, extend comparisons to additional collective libraries, and run fully controlled allocations at larger scale to better characterize worst-case contention across many inter-group links.

\section*{Acknowledgment}
\addcontentsline{toc}{section}{Acknowledgment}

This work has received funding from the European High-Performance Computing Joint Undertaking (EuroHPC JU) under grant agreement No 101175702. The project received access to the JUPITER supercomputer, which is funded by the EuroHPC JU and the German BMFTR and MKW-NRW, through the JUPITER Research and Early Access Program (JUREAP). We acknowledge the CINECA award under the ISCRA initiative, CSCS, and CSC for the availability of high-performance computing resources and support. 

\newpage

\bibliographystyle{IEEEtran} 
\bibliography{references}

\newpage

\appendix

\begin{center}
    \textit{\dlnetbench Pseudocode and Constants}
\end{center}


\begin{table}[h]
\centering
\caption{Experimental Setup and Parameters}
\label{tab:constants}
\begin{tabular}{lcccccc}
\hline
\textbf{Strategy} & \textbf{Batch Size} & \textbf{Microbatches} & \textbf{PP Degree} & \textbf{TP Degree} & \textbf{EP Degree} & \textbf{Specific Parameters} \\
\hline
DP       & 16 & -- & -- & -- & -- & \texttt{num\_buckets}=50 \\
FSDP     & 16 & -- & -- & -- & -- & \texttt{num\_units}=16, \texttt{sharding\_factor}=8 \\
DP+PP    & 16 & 16 & 8  & -- & -- & -- \\
DP+PP+TP & 16 & 16 & 8  & 4  & -- & -- \\
DP+PP+EP & 16 & 16 & 8  & -- & 8  & -- \\
\hline
\end{tabular}
\end{table}

\begin{algorithm}[h]
    \caption{\strdp: Pure Data Parallelism}
    \begin{algorithmic}[1]
        \Function{run\_data\_parallel}{}
            \State \Call{Compute}{forward time} \Comment{Forward pass}
            
            \For{each bucket}\Comment{Backward pass}
                \State \Call{Compute}{backward time per bucket}
                \State \Call{\coll{\iallreduce}{\world}}{bucket}
            \EndFor
            
            \State \Call{\coll{\waitall}{\world}}{}
        \EndFunction
    \end{algorithmic}
    \label{alg:DP}
\end{algorithm}

\begin{algorithm}[h!]
\caption{\strfsdp: Fully Sharded Data Parallelism}
\begin{algorithmic}[1]
\Function{run\_fsdp}{}

    \State \Call{\coll{\allgather}{\ir}}{unit 0}

    \For{$u = 0$ \textbf{ to } $num\_units - 2$}\Comment{Forward pass}
        \State \Call{\coll{\iallgather}{\ir}}{unit $u+1$}
        \State \Call{Compute}{forward time}
        \State \Call{\coll{\wait}{\ir}}{unit $u+1$}
    \EndFor

    \For{$u = num\_units - 1$ \textbf{ to } $0$}\Comment{Backward pass}
        \If{$u > 0$}
            \State \Call{\coll{\iallgather}{\ir}}{unit $u-1$}
        \EndIf
        \State \Call{Compute}{backward time}
        \State \Call{\coll{\reducescatter}{\ir}}{unit $u$}
        \If{num\_replicas $> 1$}
            \State \Call{\coll{\iallreduce}{\er}}{unit $u$}
        \EndIf
        \If{$u > 0$}
            \State \Call{\coll{\wait}{\ir}}{unit $u-1$}
        \EndIf
    \EndFor

    \If{num\_replicas $> 1$}
        \State \Call{\coll{\waitall}{\er}}{}
    \EndIf
\EndFunction
\end{algorithmic}
\label{alg:FSDP}
\end{algorithm}

\begin{algorithm}[h]
\caption{\strdppp: Data and Pipeline Parallelism (GPipe)}
\begin{algorithmic}[1]
\Function{run\_data\_pipeline\_parallel}{}

    \For{each microbatch}\Comment{Forward pass}
        \If{first stage}
            \State \Call{Compute}{microbatch forward time}
            \State \Call{\coll{\send}{\ir}}{next stage}
        \ElsIf{last stage}
            \State \Call{\coll{\recv}{\ir}}{previous stage}
            \State \Call{Compute}{microbatch forward time}
        \Else
            \State \Call{\coll{\recv}{\ir}}{previous stage}
            \State \Call{Compute}{microbatch forward time}
            \State \Call{\coll{\send}{\ir}}{next stage}
        \EndIf
    \EndFor

    \For{each microbatch}\Comment{Backward pass}
        \If{first stage}
            \State \Call{\coll{\recv}{\ir}}{next stage}
            \State \Call{Compute}{microbatch backward time}
        \ElsIf{last stage}
            \State \Call{Compute}{microbatch backward time}
            \State \Call{\coll{\send}{\ir}}{previous stage}
        \Else
            \State \Call{\coll{\recv}{\ir}}{next stage}
            \State \Call{Compute}{microbatch backward time}
            \State \Call{\coll{\send}{\ir}}{previous stage}
        \EndIf
    \EndFor

    \State \Call{\coll{\allreduce}{\er}}{gradients} \Comment{Data parallel gradient synchronization}
\EndFunction
\end{algorithmic}
\label{alg:GPipe}
\end{algorithm}

\begin{algorithm}
\caption{\strdppptp: Data, Pipeline and Tensor Parallelism}
\begin{algorithmic}[1]
\Function{run\_data\_pipeline\_tensor\_parallel}{}

    \For{each microbatch} \Comment{Forward pass}
        \If{first stage}
            \For{each layer}
                \State \Call{Compute}{attention forward time}
                \State \Call{\coll{\allreduce}{\ir}}{tensor parallel}
                \State \Call{Compute}{MLP forward time}
                \State \Call{\coll{\allreduce}{\ir}}{tensor parallel}
            \EndFor
            \State \Call{\coll{\send}{\ir}}{next stage}
        \ElsIf{last stage}
            \State \Call{\coll{\recv}{\ir}}{previous stage}
            \For{each layer}
                \State \Call{Compute}{attention forward time}
                \State \Call{\coll{\allreduce}{\ir}}{tensor parallel}
                \State \Call{Compute}{MLP forward time}
                \State \Call{\coll{\allreduce}{\ir}}{tensor parallel}
            \EndFor
        \Else
            \State \Call{\coll{\recv}{\ir}}{previous stage}
            \For{each layer}
                \State \Call{Compute}{attention forward time}
                \State \Call{\coll{\allreduce}{\ir}}{tensor parallel}
                \State \Call{Compute}{MLP forward time}
                \State \Call{\coll{\allreduce}{\ir}}{tensor parallel}
            \EndFor
            \State \Call{\coll{\send}{\ir}}{next stage}
        \EndIf
    \EndFor

    \For{each microbatch}\Comment{Backward pass}
        \If{first stage}
            \State \Call{\coll{\recv}{\ir}}{next stage}
            \For{each layer}
                \State \Call{Compute}{attention backward time}
                \State \Call{\coll{\allreduce}{\ir}}{tensor parallel}
                \State \Call{Compute}{MLP backward time}
                \State \Call{\coll{\allreduce}{\ir}}{tensor parallel}
            \EndFor
        \ElsIf{last stage}
            \For{each layer}
                \State \Call{Compute}{attention backward time}
                \State \Call{\coll{\allreduce}{\ir}}{tensor parallel}
                \State \Call{Compute}{MLP backward time}
                \State \Call{\coll{\allreduce}{\ir}}{tensor parallel}
            \EndFor
            \State \Call{\coll{\send}{\ir}}{previous stage}
        \Else
            \State \Call{\coll{\recv}{\ir}}{next stage}
            \For{each layer}
                \State \Call{Compute}{attention backward time}
                \State \Call{\coll{\allreduce}{\ir}}{tensor parallel}
                \State \Call{Compute}{MLP backward time}
                \State \Call{\coll{\allreduce}{\ir}}{tensor parallel}
            \EndFor
            \State \Call{\coll{\send}{\ir}}{previous stage}
        \EndIf
    \EndFor

    \State \Comment{Data parallel gradient synchronization}
    \State \Call{\coll{\allreduce}{\er}}{gradients}
\EndFunction
\end{algorithmic}
\label{alg:GPipe-TP}
\end{algorithm}

\begin{algorithm}
\caption{\strdpppep: Data, Pipeline and Expert Parallelism}
\begin{algorithmic}[1]
\Function{run\_data\_pipeline\_expert\_parallel}{}

    \For{each microbatch} \Comment{Forward pass}
        \If{first stage}
            \For{each layer}
                \State \Call{Compute}{attention forward time}
                \State \Call{\coll{\alltoall}{\ir}}{expert parallel}
                \State \Call{Compute}{MLP forward time}
                \State \Call{\coll{\alltoall}{\ir}}{expert parallel}
            \EndFor
            \State \Call{\coll{\send}{\ir}}{next stage}
        \ElsIf{last stage}
            \State \Call{\coll{\recv}{\ir}}{previous stage}
            \For{each layer}
                \State \Call{Compute}{attention forward time}
                \State \Call{\coll{\alltoall}{\ir}}{expert parallel}
                \State \Call{Compute}{MLP forward time}
                \State \Call{\coll{\alltoall}{\ir}}{expert parallel}
            \EndFor
        \Else
            \State \Call{\coll{\recv}{\ir}}{previous stage}
            \For{each layer}
                \State \Call{Compute}{attention forward time}
                \State \Call{\coll{\alltoall}{\ir}}{expert parallel}
                \State \Call{Compute}{MLP forward time}
                \State \Call{\coll{\alltoall}{\ir}}{expert parallel}
            \EndFor
            \State \Call{\coll{\send}{\ir}}{next stage}
        \EndIf
    \EndFor

    \For{each microbatch} \Comment{Backward pass}
        \If{first stage}
            \State \Call{\coll{\recv}{\ir}}{next stage}
            \For{each layer}
                \State \Call{Compute}{attention backward time}
                \State \Call{\coll{\alltoall}{\ir}}{expert parallel}
                \State \Call{Compute}{MLP backward time}
                \State \Call{\coll{\alltoall}{\ir}}{expert parallel}
            \EndFor
        \ElsIf{last stage}
            \For{each layer}
                \State \Call{Compute}{attention backward time}
                \State \Call{\coll{\alltoall}{\ir}}{expert parallel}
                \State \Call{Compute}{MLP backward time}
                \State \Call{\coll{\alltoall}{\ir}}{expert parallel}
            \EndFor
            \State \Call{\coll{\send}{\ir}}{previous stage}
        \Else
            \State \Call{\coll{\recv}{\ir}}{next stage}
            \For{each layer}
                \State \Call{Compute}{attention backward time}
                \State \Call{\coll{\alltoall}{\ir}}{expert parallel}
                \State \Call{Compute}{MLP backward time}
                \State \Call{\coll{\alltoall}{\ir}}{expert parallel}
            \EndFor
            \State \Call{\coll{\send}{\ir}}{previous stage}
        \EndIf
    \EndFor

    \State \Comment{Data parallel gradient synchronization}
    \State \Call{\coll{\allreduce}{\ir}}{non-expert gradients, expert parallel}
    \State \Call{\coll{\allreduce}{\er}}{gradients, data parallel}
\EndFunction
\end{algorithmic}
\label{alg:GPipe-EP}
\end{algorithm}

\end{document}